\magnification=\magstephalf
\newbox\SlashedBox 
\def\slashed#1{\setbox\SlashedBox=\hbox{#1}
\hbox to 0pt{\hbox to 1\wd\SlashedBox{\hfil/\hfil}\hss}{#1}}
\def\hboxtosizeof#1#2{\setbox\SlashedBox=\hbox{#1}
\hbox to 1\wd\SlashedBox{#2}}

\def\dfrac#1/#2{%
\hskip-.2em\kern .2em\raise.4ex\hbox{\the\scriptfont0 #1}\kern-.2em/%
\kern -.15em\lower.35ex\hbox{\the\scriptfont0 #2}}
\def\mathslashed#1{\setbox\SlashedBox=\hbox{$#1$}
\hbox to 0pt{\hbox to 1\wd\SlashedBox{\hfil/\hfil}\hss}#1}

\def\clap#1{\hbox to 0pt{\hss#1\hss}}

\def\ifsmall{\iffalse}  
\def\titlepagefont{}  

\def\DefineTeXgraphics{%
\special{ps::[global] /TeXgraphics { } def}}  

\def\today{\ifcase\month\or January\or February\or March\or April\or May
\or June\or July\or August\or September\or October\or November\or
December\fi\space\number\day, \number\year}
\def\eatPrefix19{}
\def\Year{\expandafter\eatPrefix\the\year}
\newcount\hours \newcount\minutes
\def\monthname{\ifcase\month\or
January\or February\or March\or April\or May\or June\or July\or
August\or September\or October\or November\or December\fi}
\def\shortmonthname{\ifcase\month\or
Jan\or Feb\or Mar\or Apr\or May\or Jun\or Jul\or
Aug\or Sep\or Oct\or Nov\or Dec\fi}

\def\TimeStamp{\hours\the\time\divide\hours by60%
\minutes -\the\time\divide\minutes by60\multiply\minutes by60%
\advance\minutes by\the\time%
${\rm \shortmonthname}\cdot\if\day<10{}0\fi\the\day\cdot\the\year%
\qquad\the\hours:\if\minutes<10{}0\fi\the\minutes$}




\def\Title#1{%
\vskip 1in{\titlefont\centerline{#1}}\vskip .5in}
 
\def\Date#1{\leftline{#1}\tenrm\supereject%
\global\hsize=\hsbody\global\hoffset=\hbodyoffset%
\footline={\hss\tenrm\folio\hss}}

\newif\ifdraftmode
\newif\ifleftlabels  

\def\nolabels{\def\wrlabeL##1{}\def\eqlabeL##1{}\def\reflabeL##1{}}
\def\writelabels{\def\wrlabeL##1{\leavevmode\vadjust{\rlap{\smash%
{\line{{\escapechar=` \hfill\rlap{\sevenrm\hskip.03in\string##1}}}}}}}%
\def\eqlabeL##1{{\escapechar-1\rlap{\sevenrm\hskip.05in\string##1}}}%
\def\reflabeL##1{\noexpand\rlap{\noexpand\sevenrm[\string##1]}}}
\def\writeleftlabels{\def\wrlabeL##1{\leavevmode\vadjust{\rlap{\smash%
{\line{{\escapechar=` \hfill\rlap{\sevenrm\hskip.03in\string##1}}}}}}}%
\def\eqlabeL##1{{\escapechar-1%
\rlap{\sixrm\hskip.05in\string##1}%
\llap{\sevenrm\string##1\hskip.03in\hbox to \hsize{}}}}%
\def\reflabeL##1{\noexpand\rlap{\noexpand\sevenrm[\string##1]}}}
\nolabels

\newdimen\fullhsize
\newdimen\hstitle
\hstitle=\hsize 
\newdimen\hsbody
\hsbody=\hsize 
\newdimen\hbodyoffset
\hbodyoffset=\hoffset 
\newbox\leftpage
\def\abstract#1{#1}
\def\rotated{\special{ps: landscape}
\magnification=1000  
\baselineskip=14pt
\global\hstitle=9truein\global\hsbody=4.75truein
\global\vsize=7truein\global\voffset=-.31truein
\global\hoffset=-0.54in\global\hbodyoffset=-.54truein
\global\fullhsize=10truein
\def\DefineTeXgraphics{%
\special{ps::[global] 
/TeXgraphics {currentpoint translate 0.7 0.7 scale
              -80 0.72 mul -1000 0.72 mul translate} def}}
\let\lr=L
\def\ifsmall{\iftrue}
\def\titlepagefont{\twelvepoint}
\trueseventeenpoint
\def\almostshipout##1{\if L\lr \count1=1
      \global\setbox\leftpage=##1 \global\let\lr=R
   \else \count1=2
      \shipout\vbox{\hbox to\fullhsize{\box\leftpage\hfil##1}}
      \global\let\lr=L\fi}

\output={\ifnum\count0=1 
 \shipout\vbox{\hbox to \fullhsize{\hfill\pagebody\hfill}}\advancepageno
 \else
 \almostshipout{\leftline{\vbox{\pagebody\makefootline}}}\advancepageno 
 \fi}

\def\abstract##1{{\leftskip=1.5in\rightskip=1.5in ##1\par}} }

\def\linemessage#1{\immediate\write16{#1}}

\global\newcount\secno \global\secno=0
\global\newcount\appno \global\appno=0
\global\newcount\meqno \global\meqno=1
\global\newcount\subsecno \global\subsecno=0
\global\newcount\figno \global\figno=0

\newif\ifAnyCounterChanged
\let\terminator=\relax
\def\normalize#1{\ifx#1\terminator\let\next=\relax\else%
\if#1i\aftergroup i\else\if#1v\aftergroup v\else\if#1x\aftergroup x%
\else\if#1l\aftergroup l\else\if#1c\aftergroup c\else%
\if#1m\aftergroup m\else%
\if#1I\aftergroup I\else\if#1V\aftergroup V\else\if#1X\aftergroup X%
\else\if#1L\aftergroup L\else\if#1C\aftergroup C\else%
\if#1M\aftergroup M\else\aftergroup#1\fi\fi\fi\fi\fi\fi\fi\fi\fi\fi\fi\fi%
\let\next=\normalize\fi%
\next}
\def\makeNormal#1#2{\def\doNormalDef{\edef#1}\begingroup%
\aftergroup\doNormalDef\aftergroup{\normalize#2\terminator\aftergroup}%
\endgroup}

\def\warnIfChanged#1#2{%
\ifundef#1
\else\begingroup%
\edef\oldDefinitionOfCounter{#1}\edef\newDefinitionOfCounter{#2}%
\ifx\oldDefinitionOfCounter\newDefinitionOfCounter%
\else%
\linemessage{Warning: definition of \noexpand#1 has changed.}%
\global\AnyCounterChangedtrue\fi\endgroup\fi}

\def\Section#1{\global\advance\secno by1\relax\global\meqno=1%
\global\subsecno=0%
\bigbreak\bigskip
\centerline{\twelvepoint \bf %
\the\secno. #1}%
\par\nobreak\medskip\nobreak}
\def\tagsection#1{%
\warnIfChanged#1{\the\secno}%
\xdef#1{\the\secno}%
\ifWritingAuxFile\immediate\write\auxfile{\noexpand\xdef\noexpand#1{#1}}\fi%
}
\def\section{\Section}
\def\Subsection#1{\global\advance\subsecno by1\relax\medskip %
\leftline{\bf\the\secno.\the\subsecno\ #1}%
\par\nobreak\smallskip\nobreak}
\def\tagsubsection#1{%
\warnIfChanged#1{\the\secno.\the\subsecno}%
\xdef#1{\the\secno.\the\subsecno}%
\ifWritingAuxFile\immediate\write\auxfile{\noexpand\xdef\noexpand#1{#1}}\fi%
}

\def\subsection{\Subsection}

\def\romappno{\uppercase\expandafter{\romannumeral\appno}}
\def\makeNormalizedRomappno{%
\expandafter\makeNormal\expandafter\normalizedromappno%
\expandafter{\romannumeral\appno}%
\edef\normalizedromappno{\uppercase{\normalizedromappno}}}
\def\Appendix#1{\global\advance\appno by1\relax\global\meqno=1\global\secno=0%
\global\subsecno=0%
\bigbreak\bigskip
\centerline{\twelvepoint \bf Appendix %
\romappno. #1}%
\par\nobreak\medskip\nobreak}
\def\tagappendix#1{\makeNormalizedRomappno%
\warnIfChanged#1{\normalizedromappno}%
\xdef#1{\normalizedromappno}%
\ifWritingAuxFile\immediate\write\auxfile{\noexpand\xdef\noexpand#1{#1}}\fi%
}
\def\appendix{\Appendix}
\def\Subappendix#1{\global\advance\subsecno by1\relax\medskip %
\leftline{\bf\romappno.\the\subsecno\ #1}%
\par\nobreak\smallskip\nobreak}
\def\tagsubappendix#1{\makeNormalizedRomappno%
\warnIfChanged#1{\normalizedromappno.\the\subsecno}%
\xdef#1{\normalizedromappno.\the\subsecno}%
\ifWritingAuxFile\immediate\write\auxfile{\noexpand\xdef\noexpand#1{#1}}\fi%
}

\def\eqn#1{\makeNormalizedRomappno%
\ifnum\secno>0%
  \warnIfChanged#1{\the\secno.\the\meqno}%
  \eqno(\the\secno.\the\meqno)\xdef#1{\the\secno.\the\meqno}%
     \global\advance\meqno by1
\else\ifnum\appno>0%
  \warnIfChanged#1{\normalizedromappno.\the\meqno}%
  \eqno({\rm\romappno}.\the\meqno)%
      \xdef#1{\normalizedromappno.\the\meqno}%
     \global\advance\meqno by1
\else%
  \warnIfChanged#1{\the\meqno}%
  \eqno(\the\meqno)\xdef#1{\the\meqno}%
     \global\advance\meqno by1
\fi\fi%
\eqlabeL#1%
\ifWritingAuxFile\immediate\write\auxfile{\noexpand\xdef\noexpand#1{#1}}\fi%
}
\def\defeqn#1{\makeNormalizedRomappno%
\ifnum\secno>0%
  \warnIfChanged#1{\the\secno.\the\meqno}%
  \xdef#1{\the\secno.\the\meqno}%
     \global\advance\meqno by1
\else\ifnum\appno>0%
  \warnIfChanged#1{\normalizedromappno.\the\meqno}%
  \xdef#1{\normalizedromappno.\the\meqno}%
     \global\advance\meqno by1
\else%
  \warnIfChanged#1{\the\meqno}%
  \xdef#1{\the\meqno}%
     \global\advance\meqno by1
\fi\fi%
\eqlabeL#1%
\ifWritingAuxFile\immediate\write\auxfile{\noexpand\xdef\noexpand#1{#1}}\fi%
}
\def\anoneqn{\makeNormalizedRomappno%
\ifnum\secno>0
  \eqno(\the\secno.\the\meqno)%
     \global\advance\meqno by1
\else\ifnum\appno>0
  \eqno({\rm\normalizedromappno}.\the\meqno)%
     \global\advance\meqno by1
\else
  \eqno(\the\meqno)%
     \global\advance\meqno by1
\fi\fi%
}
\def\mfig#1#2{\ifx#20
\else\global\advance\figno by1%
\relax#1\the\figno%
\warnIfChanged#2{\the\figno}%
\xdef#2{\the\figno}%
\reflabeL#2%
\ifWritingAuxFile\immediate\write\auxfile{\noexpand\xdef\noexpand#2{#2}}\fi\fi%
}

\catcode`@=11 

\newif\ifFiguresInText\FiguresInTexttrue
\newif\if@FigureFileCreated
\newwrite\capfile
\newwrite\figfile

\newif\ifcaption
\captiontrue
\def\captionsize{\tenrm}
\def\PlaceTextFigure#1#2#3#4{%
\vskip 0.5truein%
\noindent#3\hfil\epsfbox{#4}\hfil\break%
\ifcaption\vskip 5pt\noindent\hfil\vbox{\captionsize \noindent Figure #1. #2}\hfil\fi%
\vskip10pt}
\def\PlaceEndFigure#1#2{%
\epsfxsize=\hsize\epsfbox{#2}\vfill\centerline{Figure #1.}\eject}

\def\LoadFigure#1#2#3#4{%
\vphantom{\mfig{}#1}
\ifx#10
\else
\fi
\ifFiguresInText
\PlaceTextFigure{#1}{#2}{#3}{#4}%
\else
\if@FigureFileCreated\else%
\immediate\openout\capfile=\jobname.caps%
\immediate\openout\figfile=\jobname.figs%
@FigureFileCreatedtrue\fi%
\immediate\write\capfile{\noexpand\item{Figure \noexpand#1.\ }{#2}\vskip10pt}%
\immed	iate\write\figfile{\noexpand\PlaceEndFigure\noexpand#1{\noexpand#4}}%
\fi}

\def\listfigs{\ifFiguresInText\else%
\vfill\eject\immediate\closeout\capfile
\immediate\closeout\figfile%
\centerline{{\bf Figures}}\bigskip\frenchspacing%
\catcode`@=11 
\def\captionsize{\tenrm}
\input \jobname.caps\vfill\eject\nonfrenchspacing%
\catcode`\@=\active
\catcode`@=12  
\input\jobname.figs\fi}

\font\ninerm=cmr9
\font\eightrm=cmr8
\font\sixrm=cmr6

\def\loadtrueseventeenpoint{
 \font\seventeenrm=cmr10 at 17.28truept
 \font\seventeeni=cmmi10 at 17.28truept
 \font\seventeenbf=cmbx10 at 17.28truept
 \font\seventeenit=cmti10 at 17.28truept
 \font\seventeensl=cmsl10 at 17.28truept
 \font\seventeensy=cmsy10 at 17.28truept
}
\def\loadfourteenpoint{
\font\fourteenrm=cmr10 at 14.4pt
\font\fourteeni=cmmi10 at 14.4pt
\font\fourteenit=cmti10 at 14.4pt
\font\fourteensl=cmsl10 at 14.4pt
\font\fourteensy=cmsy10 at 14.4pt
\font\fourteenbf=cmbx10 at 14.4pt
}
\def\loadtruetwelvepoint{
\font\twelverm=cmr10 at 12truept
\font\twelvei=cmmi10 at 12truept
\font\twelveit=cmti10 at 12truept
\font\twelvesl=cmsl10 at 12truept
\font\twelvesy=cmsy10 at 12truept
\font\twelvebf=cmbx10 at 12truept
\font\twelvesc=cmcsc10 at 12truept
}

\font\ninei=cmmi9
\font\eighti=cmmi8
\font\sixi=cmmi6
\skewchar\ninei='177 \skewchar\eighti='177 \skewchar\sixi='177

\font\ninesy=cmsy9
\font\eightsy=cmsy8
\font\sixsy=cmsy6
\skewchar\ninesy='60 \skewchar\eightsy='60 \skewchar\sixsy='60

\font\ninebf=cmbx9
\font\eightbf=cmbx8
\font\sixbf=cmbx6

\font\ninett=cmtt9
\font\eighttt=cmtt8

\hyphenchar\tentt=-1 
\hyphenchar\ninett=-1
\hyphenchar\eighttt=-1         

\font\ninesl=cmsl9
\font\eightsl=cmsl8

\font\nineit=cmti9
\font\eightit=cmti8
\font\sevenit=cmti7

\scriptfont\itfam=\sevenit


                      
\newskip\ttglue
\def\tenpoint{\def\rm{\fam0\tenrm}%
  \textfont0=\tenrm \scriptfont0=\sevenrm \scriptscriptfont0=\fiverm
  \textfont1=\teni \scriptfont1=\seveni \scriptscriptfont1=\fivei
  \textfont2=\tensy \scriptfont2=\sevensy \scriptscriptfont2=\fivesy
  \textfont3=\tenex \scriptfont3=\tenex \scriptscriptfont3=\tenex
  \def\it{\fam\itfam\tenit}%
      \textfont\itfam=\tenit\scriptfont\itfam=\sevenit
  \def\sl{\fam\slfam\tensl}\textfont\slfam=\tensl
  \def\bf{\fam\bffam\tenbf}\textfont\bffam=\tenbf \scriptfont\bffam=\sevenbf
  \scriptscriptfont\bffam=\fivebf
  \normalbaselineskip=12pt
  \let\sc=\eightrm
  \let\big=\tenbig
  \setbox\strutbox=\hbox{\vrule height8.5pt depth3.5pt width\z@}%
  \normalbaselines\rm}

\def\twelvepoint{\def\rm{\fam0\twelverm}%
  \textfont0=\twelverm \scriptfont0=\ninerm \scriptscriptfont0=\sevenrm
  \textfont1=\twelvei \scriptfont1=\ninei \scriptscriptfont1=\seveni
  \textfont2=\twelvesy \scriptfont2=\ninesy \scriptscriptfont2=\sevensy
  \textfont3=\tenex \scriptfont3=\tenex \scriptscriptfont3=\tenex
  \def\it{\fam\itfam\twelveit}\textfont\itfam=\twelveit
  \def\sl{\fam\slfam\twelvesl}\textfont\slfam=\twelvesl
  \def\bf{\fam\bffam\twelvebf}\textfont\bffam=\twelvebf%
  \scriptfont\bffam=\ninebf
  \scriptscriptfont\bffam=\sevenbf
  \normalbaselineskip=12pt
  \let\sc=\eightrm
  \let\big=\tenbig
  \setbox\strutbox=\hbox{\vrule height8.5pt depth3.5pt width\z@}%
  \normalbaselines\rm}

\def\fourteenpoint{\def\rm{\fam0\fourteenrm}%
  \textfont0=\fourteenrm \scriptfont0=\tenrm \scriptscriptfont0=\sevenrm
  \textfont1=\fourteeni \scriptfont1=\teni \scriptscriptfont1=\seveni
  \textfont2=\fourteensy \scriptfont2=\tensy \scriptscriptfont2=\sevensy
  \textfont3=\tenex \scriptfont3=\tenex \scriptscriptfont3=\tenex
  \def\it{\fam\itfam\fourteenit}\textfont\itfam=\fourteenit
  \def\sl{\fam\slfam\fourteensl}\textfont\slfam=\fourteensl
  \def\bf{\fam\bffam\fourteenbf}\textfont\bffam=\fourteenbf%
  \scriptfont\bffam=\tenbf
  \scriptscriptfont\bffam=\sevenbf
  \normalbaselineskip=17pt
  \let\sc=\elevenrm
  \let\big=\tenbig                                          
  \setbox\strutbox=\hbox{\vrule height8.5pt depth3.5pt width\z@}%
  \normalbaselines\rm}

\def\seventeenpoint{\def\rm{\fam0\seventeenrm}%
  \textfont0=\seventeenrm \scriptfont0=\fourteenrm \scriptscriptfont0=\tenrm
  \textfont1=\seventeeni \scriptfont1=\fourteeni \scriptscriptfont1=\teni
  \textfont2=\seventeensy \scriptfont2=\fourteensy \scriptscriptfont2=\tensy
  \textfont3=\tenex \scriptfont3=\tenex \scriptscriptfont3=\tenex
  \def\it{\fam\itfam\seventeenit}\textfont\itfam=\seventeenit
  \def\sl{\fam\slfam\seventeensl}\textfont\slfam=\seventeensl
  \def\bf{\fam\bffam\seventeenbf}\textfont\bffam=\seventeenbf%
  \scriptfont\bffam=\fourteenbf
  \scriptscriptfont\bffam=\twelvebf
  \normalbaselineskip=21pt
  \let\sc=\fourteenrm
  \let\big=\tenbig                                          
  \setbox\strutbox=\hbox{\vrule height 12pt depth 6pt width\z@}%
  \normalbaselines\rm}

\def\ninepoint{\def\rm{\fam0\ninerm}%
  \textfont0=\ninerm \scriptfont0=\sixrm \scriptscriptfont0=\fiverm
  \textfont1=\ninei \scriptfont1=\sixi \scriptscriptfont1=\fivei
  \textfont2=\ninesy \scriptfont2=\sixsy \scriptscriptfont2=\fivesy
  \textfont3=\tenex \scriptfont3=\tenex \scriptscriptfont3=\tenex
  \def\it{\fam\itfam\nineit}\textfont\itfam=\nineit
  \def\sl{\fam\slfam\ninesl}\textfont\slfam=\ninesl
  \def\bf{\fam\bffam\ninebf}\textfont\bffam=\ninebf \scriptfont\bffam=\sixbf
  \scriptscriptfont\bffam=\fivebf
  \normalbaselineskip=11pt
  \let\sc=\sevenrm
  \let\big=\ninebig
  \setbox\strutbox=\hbox{\vrule height8pt depth3pt width\z@}%
  \normalbaselines\rm}

\def\eightpoint{\def\rm{\fam0\eightrm}%
  \textfont0=\eightrm \scriptfont0=\sixrm \scriptscriptfont0=\fiverm%
  \textfont1=\eighti \scriptfont1=\sixi \scriptscriptfont1=\fivei%
  \textfont2=\eightsy \scriptfont2=\sixsy \scriptscriptfont2=\fivesy%
  \textfont3=\tenex \scriptfont3=\tenex \scriptscriptfont3=\tenex%
  \def\it{\fam\itfam\eightit}\textfont\itfam=\eightit%
  \def\sl{\fam\slfam\eightsl}\textfont\slfam=\eightsl%
  \def\bf{\fam\bffam\eightbf}\textfont\bffam=\eightbf \scriptfont\bffam=\sixbf%
  \scriptscriptfont\bffam=\fivebf%
  \normalbaselineskip=9pt%
  \let\sc=\sixrm%
  \let\big=\eightbig%
  \setbox\strutbox=\hbox{\vrule height7pt depth2pt width\z@}%
  \normalbaselines\rm}
  \let\sc=\eightrm

\def\tenbig#1{{\hbox{$\left#1\vbox to8.5pt{}\right.\n@space$}}}
\def\ninebig#1{{\hbox{$\textfont0=\tenrm\textfont2=\tensy
  \left#1\vbox to7.25pt{}\right.\n@space$}}}
\def\eightbig#1{{\hbox{$\textfont0=\ninerm\textfont2=\ninesy
  \left#1\vbox to6.5pt{}\right.\n@space$}}}

\def\footnote#1{\edef\@sf{\spacefactor\the\spacefactor}#1\@sf
      \insert\footins\bgroup\eightpoint
      \interlinepenalty100 \let\par=\endgraf
        \leftskip=\z@skip \rightskip=\z@skip
        \splittopskip=10pt plus 1pt minus 1pt \floatingpenalty=20000
        \smallskip\item{#1}\bgroup\strut\aftergroup\@foot\let\next}
\skip\footins=12pt plus 2pt minus 4pt 
\dimen\footins=30pc 

\newinsert\margin
\dimen\margin=\maxdimen
\def\titlefont{\seventeenpoint}
\loadtruetwelvepoint 
\loadtrueseventeenpoint

\def\eatOne#1{}
\def\ifundef#1{\expandafter\ifx%
\csname\expandafter\eatOne\string#1\endcsname\relax}
\def\notTrue{\iffalse}\def\isTrue{\iftrue}
\def\ifdef#1{{\ifundef#1%
\aftergroup\notTrue\else\aftergroup\isTrue\fi}}
\def\use#1{\ifundef#1\linemessage{Warning: \string#1 is undefined.}%
{\tt \string#1}\else#1\fi}



%
\catcode`"=11
\let\quote="
\catcode`"=12
\chardef\foo="22
\global\newcount\refno \global\refno=1
\newwrite\rfile
\newlinechar=`\^^J
\def\@ref#1#2{\the\refno\n@ref#1{#2}}
\def\h@ref#1#2#3{\href{#3}{\the\refno}\n@ref#1{#2}}
\def\n@ref#1#2{\xdef#1{\the\refno}%
\ifnum\refno=1\immediate\openout\rfile=\jobname.refs\fi%
\immediate\write\rfile{\noexpand\item{[\noexpand#1]\ }#2.}%
\global\advance\refno by1}
\def\nref{\n@ref} 
\def\ref{\@ref}   
\def\hrref{\h@ref}
\def\lref#1#2{\the\refno\xdef#1{\the\refno}%
\ifnum\refno=1\immediate\openout\rfile=\jobname.refs\fi%
\immediate\write\rfile{\noexpand\item{[\noexpand#1]\ }#2\semi}%
\global\advance\refno by1}
\def\cref#1{\immediate\write\rfile{#1\semi}}

\def\preref#1#2{\gdef#1{\@ref#1{#2}}}

\def\semi{;\hfil\noexpand\break}

\def\listrefs{\vfill\eject\immediate\closeout\rfile
\centerline{{\bf References}}\bigskip\frenchspacing%
\input \jobname.refs\vfill\eject\nonfrenchspacing}

\def\inputAuxIfPresent#1{\immediate\openin1=#1
\ifeof1\message{No file \auxfileName; I'll create one.
}\else\closein1\relax\input\auxfileName\fi%
}
\def\NPB{Nucl.\ Phys.\ B}




\newif\ifWritingAuxFile
\newwrite\auxfile
\def\SetUpAuxFile{%
\xdef\auxfileName{\jobname.aux}%
\inputAuxIfPresent{\auxfileName}%
\WritingAuxFiletrue%
\immediate\openout\auxfile=\auxfileName}


\def\bye{\par\vfill\supereject%
\ifAnyCounterChanged\linemessage{
Some counters have changed.  Re-run tex to fix them up.}\fi%
\end}

\catcode`\@=\active
\catcode`@=12  
\catcode`\"=\active

\def\mod{\mathop{\rm mod}\nolimits}

\def\Ksl{\slashed{K}}

\def\spa#1.#2{\left\langle#1\,#2\right\rangle}
\def\spb#1.#2{\left[#1\,#2\right]}
\def\lor#1.#2{\left(#1\,#2\right)}
\def\sand#1.#2.#3{%
\left\langle\smash{#1}{\vphantom1}^{-}\right|{#2}%
\left|\smash{#3}{\vphantom1}^{-}\right\rangle}
\def\sandp#1.#2.#3{%
\left\langle\smash{#1}{\vphantom1}^{-}\right|{#2}%
\left|\smash{#3}{\vphantom1}^{+}\right\rangle}
\def\sandpp#1.#2.#3{%
\left\langle\smash{#1}{\vphantom1}^{+}\right|{#2}%
\left|\smash{#3}{\vphantom1}^{+}\right\rangle}
\def\sandpm#1.#2.#3{%
\left\langle\smash{#1}{\vphantom1}^{+}\right|{#2}%
\left|\smash{#3}{\vphantom1}^{-}\right\rangle}
\def\sandmp#1.#2.#3{%
\left\langle\smash{#1}{\vphantom1}^{-}\right|{#2}%
\left|\smash{#3}{\vphantom1}^{+}\right\rangle}
\catcode`@=11  
\def\meqalign#1{\,\vcenter{\openup1\jot\m@th
   \ialign{\strut\hfil$\displaystyle{##}$ && $\displaystyle{{}##}$\hfil
             \crcr#1\crcr}}\,}
\catcode`@=12  

\input epsf
\def\captionsize{\ninerm}
\font\tenmbb=msbm10
\font\sevenmbb=msbm7
\newfam\mbbfam
\def\mathbb{\fam\mbbfam\tenmbb}\textfont\mbbfam=\tenmbb
\scriptfont\mbbfam=\sevenmbb

\SetUpAuxFile
\hfuzz 20pt
\overfullrule 0pt

\def\e{\epsilon}
\def\tree{{\rm tree\vphantom{p}}}
\def\dash{\hbox{-\kern-.02em}}
\def\oneloop{{\rm 1\dash{}loop}}

\def\dash{\hbox{-\kern-.02em}}

\def\llongrightarrow{%
\relbar\mskip-0.5mu\joinrel\mskip-0.5mu\relbar\mskip-0.5mu\joinrel\longrightarrow}
\def\inlimit^#1{\buildrel#1\over\llongrightarrow}
\def\frac#1#2{{#1\over #2}}

\def\la{\lu{a}}
\def\lb{\lu{b}}

\def\proj{\flat}

\def\tspa#1.#2{\left\langle#1,#2\right\rangle}
\def\tspb#1.#2{\left[#1,#2\right]}
\def\Li{\mathop{\rm Li}\nolimits}

\preref\DixonTASI{L.\ Dixon, in 
{\it QCD \& Beyond: Proceedings of TASI '95}, 
ed. D.\ E.\ Soper (World Scientific, 1996) [hep-ph/9601359]}
\preref\Recurrence{F.\ A.\ Berends and W.\ T.\ Giele, 
Nucl.\ Phys.\ B306:759 (1988)}
\preref\HelicityRecurrence{
D.\ A.\ Kosower,
Nucl.\ Phys.\ B335:23 (1990)%
}
\preref\BernChalmers{
Z.\ Bern and G.\ Chalmers,
Nucl.\ Phys.\ B447:465 (1995) [hep-ph/9503236]%
}
\preref\ManganoParke{M.\ Mangano and S.\ J.\ Parke, Phys.\ Rep.\ 200:301 (1991)}
\preref\SingleAntenna{
D.\ A.\ Kosower,
Phys.\ Rev.\ D57:5410 (1998) [hep-ph/9710213]%
}
\preref\MultipleAntenna{
D.\ A.\ Kosower,
Phys.\ Rev.\ D67:116003 (2003) [hep-ph/0212097]%
}
\preref\NeqFourOneLoop{Z. Bern, L. J. Dixon, D. C. Dunbar, and D. A. Kosower,
Nucl.\ Phys.\ B425:217 (1994) [hep-ph/9403226]}
\preref\AllOrdersCollinear{
D. A. Kosower,
Nucl.\ Phys.\ B552:319 (1999) [hep-ph/9901201]%
}
\preref\DDFM{
V.\ Del Duca, A.\ Frizzo and F.\ Maltoni,
Nucl.\ Phys.\ B568:211 (2000) [hep-ph/9909464]%
}
\preref\GloverCampbell{
J.\ M.\ Campbell and E.\ W.\ N.\ Glover,
Nucl.\ Phys.\ B527:264 (1998) [hep-ph/9710255]%
}
\preref\CataniGrazzini{
S.\ Catani and M.\ Grazzini,
Phys.\ Lett.\ B446:143 (1999) [hep-ph/9810389]\semi
S.\ Catani and M.\ Grazzini,
Nucl.\ Phys.\ B570:287 (2000) [hep-ph/9908523]%
}
\preref\BerendsGieleSoft{
F.\ A.\ Berends and W.\ T.\ Giele,
Nucl.\ Phys.\ B313:595 (1989)%
}
\preref\AltarelliParisi{G.\ Altarelli and G.\ Parisi, Nucl.\ Phys.\ B126:298 (1977)}
\preref\GieleGlover{W.\ T.\ Giele and E.\ W.\ N.\ Glover, 
Phys.\ Rev.\ D46:1980 (1992)}
\preref\GieleGloverKosower{
W.\ T.\ Giele, E.\ W.\ N.\ Glover and D.\ A.\ Kosower,
Nucl.\ Phys.\ B403:633 (1993) [hep-ph/9302225]%
}
\preref\CataniSeymour{
S.\ Catani and M.\ H.\ Seymour,
Phys.\ Lett.\ B378:287 (1996) [hep-ph/9602277]\semi
S.\ Catani and M.\ H.\ Seymour,
Nucl.\ Phys.\ B485:291 (1997); erratum-ibid.\ B510:503 (1997) [hep-ph/9605323]%
}
\preref\Color{%
F.\ A.\ Berends and W.\ T.\ Giele,
Nucl.\ Phys.\ B294:700 (1987)\semi
D.\ A.\ Kosower, B.-H.\ Lee and V.\ P.\ Nair, Phys.\ Lett.\ 201B:85 (1988)\semi
M.\ Mangano, S.\ Parke and Z.\ Xu, Nucl.\ Phys.\ B298:653 (1988)\semi
Z.\ Bern and D.\ A.\ Kosower, Nucl.\ Phys.\ B362:389 (1991)}
\preref\qqggg{Z. Bern, L. Dixon, and D. A. Kosower,
Nucl.\ Phys.\  B437:259 (1995) [hep-ph/9409393]}
\preref\AlternateColorDecomposition{
V.\ Del Duca, L.\ J.\ Dixon and F.\ Maltoni,
Nucl.\ Phys.\ B571: 51 (2000) [hep-ph/9910563]%
}

\preref\RecurrenceRelations{
F.\ A.\ Berends and W.\ T.\ Giele,
Nucl.\ Phys.\ B306:759 (1988)%
}

\preref\Spinor{%
F.\ A.\ Berends, R.\ Kleiss, P.\ De Causmaecker, R.\ Gastmans, and T.\ T.\ Wu,
        Phys.\ Lett.\ 103B:124 (1981)\semi
P.\ De Causmaeker, R.\ Gastmans,  W.\ Troost, and  T.\ T.\ Wu,
Nucl.\ Phys.\ B206:53 (1982)\semi
Z.\ Xu, D.-H.\ Zhang, L.\ Chang, Tsinghua University
                  preprint TUTP--84/3 (1984), unpublished\semi
R.\ Kleiss and W.\ J.\ Stirling, 
   Nucl.\ Phys.\ B262:235 (1985)\semi
   J.\ F.\ Gunion and Z.\ Kunszt, Phys.\ Lett.\ 161B:333 (1985)\semi
Z.\ Xu, D.-H.\ Zhang, and L.\ Chang, Nucl.\ Phys.\ B291:392 (1987)}

\preref\NeqOneOneLoop{
Z.\ Bern, L.\ J.\ Dixon, D.\ C.\ Dunbar and D.\ A.\ Kosower,
Nucl.\ Phys.\ B435:59 (1995) [hep-ph/9409265]
}

\preref\UnitarityMachinery{
Z.\ Bern, L.\ J.\ Dixon and D.\ A.\ Kosower,
Nucl.\ Phys.\ Proc.\ Suppl.\  51C:243 (1996)
[hep-ph/9606378]\semi
Z.\ Bern, L.\ J.\ Dixon and D.\ A.\ Kosower,
Ann.\ Rev.\ Nucl.\ Part.\ Sci.\  46:109 (1996)
[hep-ph/9602280]\semi
Z.\ Bern and A.\ G.\ Morgan,
Nucl.\ Phys.\ B467:479 (1996) [hep-ph/9511336]\semi%
Z.\ Bern, L.\ J.\ Dixon and D.\ A.\ Kosower,
JHEP 0408:012 (2004) [hep-ph/0404293]%
}

\preref\UnitarityB{%
Z.\ Bern, L.\ J.\ Dixon, D.\ C.\ Dunbar and D.\ A.\ Kosower,
Nucl.\ Phys.\ B435:59 (1995) [hep-ph/9409265]%
}
\preref\UnitarityReview{
}

\preref\UnitarityCalculations{
Z.\ Bern, L.\ J.\ Dixon, D.\ C.\ Dunbar and D.\ A.\ Kosower,
Nucl.\ Phys.\ B{425}:217 (1994)
[hep-ph/9403226]\semi
Z.\ Bern, L.\ J.\ Dixon and D.\ A.\ Kosower,
Nucl.\ Phys.\ B{437}:259 (1995)
[hep-ph/9409393]\semi
Z.\ Bern, L.\ J.\ Dixon, D.\ A.\ Kosower and S.\ Weinzierl,
Nucl.\ Phys.\ B{489}:3 (1997)
[hep-ph/9610370]\semi
Z.\ Bern, L.\ J.\ Dixon and D.\ A.\ Kosower,
Nucl.\ Phys.\ B{513}:3 (1998)
[hep-ph/9708239]\semi
Z.\ Bern, L.\ J.\ Dixon and D.\ A.\ Kosower,
JHEP {0001}:027 (2000)
[hep-ph/0001001]\semi
Z.\ Bern, A.\ De Freitas and L.\ J.\ Dixon,
JHEP {0109}:037 (2001)
[hep-ph/0109078]\semi
Z.\ Bern, A.\ De Freitas and L.\ J.\ Dixon,
JHEP {0203}:018 (2002)
[hep-ph/0201161]\semi
Z.\ Bern, A.\ De Freitas and L.\ J.\ Dixon,
JHEP {0306}:028 (2003)
[hep-ph/0304168]\semi
C.\ Anastasiou, Z.\ Bern, L.\ J.\ Dixon and D.\ A.\ Kosower,
Phys.\ Rev.\ Lett.\  91:251602 (2003)
[hep-th/0309040]%
}

\preref\HigherLoopAntenna{
D.\ A.\ Kosower,
hep-ph/0301069%
}
\preref\BernMorgan{
Z.\ Bern and A.\ G.\ Morgan,
Nucl.\ Phys.\ B467:479 (1996) [hep-ph/9511336]%
}

\preref\CataniConjecture{
S.\ Catani,
Phys.\ Lett.\ B427:161 (1998) [hep-ph/9802439]%
}
\preref\StermanTejeda{
G.\ Sterman and M.\ E.\ Tejeda-Yeomans,
Phys.\ Lett.\ B552:48 (2003) [hep-ph/0210130]%
}
\preref\vanNeerven{W.\ L.\ van\ Neerven, \NPB 268:453 (1986)}
\preref\CataniGrazziniSoft{
S.\ Catani and M.\ Grazzini,
Nucl.\ Phys.\ B591:435 (2000) [hep-ph/0007142]%
}
\preref\FKS{
S.\ Frixione, Z.\ Kunszt and A.\ Signer,
Nucl.\ Phys.\ B467:399 (1996) [hep-ph/9512328]
}
\preref\Byckling{E. Byckling and K. Kajantie, {\it Particle Kinematics\/}
(Wiley, 1973)}
\preref\OneloopSplitB{
Z.\ Bern, V.\ Del Duca and C.\ R.\ Schmidt,
Phys.\ Lett.\ B445:168 (1998) [hep-ph/9810409]\semi
Z.\ Bern, V.\ Del Duca, W.\ B.\ Kilgore and C.\ R.\ Schmidt,
Phys.\ Rev.\ D60:116001 (1999) [hep-ph/9903516]
}
\preref\OneloopSplitA{
D.\ A.\ Kosower and P.\ Uwer,
Nucl.\ Phys.\ B563:477 (1999) [hep-ph/9903515]
}

\preref\ZFourPartons{
Z.\ Bern, L.\ J.\ Dixon and D.\ A.\ Kosower,
Nucl.\ Phys.\ B513:3 (1998)
[hep-ph/9708239]%
}

\preref\SWI{M. T. Grisaru, H. N. Pendleton and P. van
Nieuwenhuizen,
Phys. Rev. D15:996 (1977)\semi
M. T. Grisaru and H.N. Pendleton, Nucl. Phys. B124:81 (1977)}
\preref\UseSWI{
S. J. Parke and T. Taylor, Phys. Lett. B157:81 (1985)\semi
Z. Kunszt, Nucl. Phys. B271:333 (1986)}
\preref\Lewin{L. Lewin, {\it Polylogarithms and associated functions\/}
(North-Holland,1981)}

\preref\ParkeTaylor{S. J. Parke and T. R. Taylor,
Phys.\ Rev.\ Lett.\ 56:2459 (1986)}

\preref\Nair{%
V.\ P.\ Nair,
Phys.\ Lett.\ B214:215 (1988)%
}

\preref\WittenTopologicalString{
E.\ Witten,
hep-th/0312171%
}

\preref\CSW{
F.\ Cachazo, P.\ Svr\v cek and E.\ Witten,
hep-th/0403047%
}

\preref\RSV{
R.\ Roiban, M.\ Spradlin and A.\ Volovich,
JHEP {0404}:012 (2004) [hep-th/0402016]\semi
R.\ Roiban and A.\ Volovich,
hep-th/0402121\semi
R.\ Roiban, M.\ Spradlin and A.\ Volovich,
hep-th/0403190%
}

\preref\Gukov{
S.\ Gukov, L.\ Motl and A.\ Neitzke,
hep-th/0404085%
}

\preref\Berkovits{
N.\ Berkovits,
hep-th/0402045\semi
N.\ Berkovits and L.\ Motl,
JHEP {0404}:056 (2004)
[hep-th/0403187]%
}

\preref\BerkovitsWitten{
N.\ Berkovits and E.\ Witten,
hep-th/0406051%
}

\preref\Vafa{
A.\ Neitzke and C.\ Vafa,
hep-th/0402128%
}

\preref\Siegel{
W.\ Siegel,
hep-th/0404255%
}

\preref\Khoze{
G.\ Georgiou and V.\ V.\ Khoze,
hep-th/0404072%
}

\preref\OtherGoogly{
C.\ J.\ Zhu,
JHEP {0404}:032 (2004)
[hep-th/0403115]\semi%
J.\ B.\ Wu and C.\ J.\ Zhu,
hep-th/0406085.
}

\preref\Popov{
A.\ D.\ Popov and C.\ Saemann,
hep-th/0405123%
}

\preref\NMHVPrivate{
L.\ J.\ Dixon, private communication}

\preref\SevenPoint{
Z.\ Bern, V.\ Del Duca, L.\ J.\ Dixon and D.\ A.\ Kosower, in progress}

\preref\BST{
A.\ Brandhuber, B.\ Spence and G.\ Travaglini,
hep-th/0407214%
}

\preref\CSWApplications{
G.\ Georgiou and V.\ V.\ Khoze,
JHEP 0405:070 (2004)
[hep-th/0404072]\semi
J.\ B.\ Wu and C.\ J.\ Zhu,
JHEP 0407:032 (2004)
[hep-th/0406085]\semi
J.\ B.\ Wu and C.\ J.\ Zhu,
hep-th/0406146\semi
G.\ Georgiou, E.\ W.\ N.\ Glover and V.\ V.\ Khoze,
JHEP 0407:048 (2004)
[hep-th/0407027]\semi
V.\ V.\ Khoze,
hep-th/0408233\semi
M.\ Lou and C.\ Wen, hep-th/0410045
}

\preref\CSWLoop{
F.\ Cachazo, P.\ Svr\v cek and E.\ Witten,
hep-th/0406177%
}
\preref\RecursiveTwistor{
I.\ Bena, Z.\ Bern and D.\ A.\ Kosower,
hep-th/0406133%
}

\preref\NMHV{
D.\ A.\ Kosower,
hep-th/0406175%
}

\preref\NairEffectiveTheory{
Y.\ Abe, V.\ P.\ Nair and M.\ I.\ Park,
hep-th/0408191%
}

\preref\CSWLoopII{
F.\ Cachazo, P.\ Svr\v cek and E.\ Witten,
hep-th/0409245%
}

\preref\DAKir{D.\ A.\ Kosower, unpublished}

\preref\KSTir{
Z.\ Kunszt, A.\ Signer and Z.\ Trocsanyi,
Nucl.\ Phys.\ B420:550 (1994) [hep-ph/9401294]%
}

\preref\Erdelyi{
A. Erd\'elyi et.\ al., {\it Tables of Integral Transforms}, Vol I, 
 (McGraw-Hill, 1954)}

\loadfourteenpoint
\noindent\nopagenumbers
[hep-th/0410054] 
\hfill\hbox to 1.5truein{\vtop{\noindent
       UCLA/04/TEP/42\break SPhT-T04/125\break PUPT-2137\break
                 NSF-KITP-04-112}\hss}

\leftlabelstrue
\vskip -0.7 in
\Title{Loops in Twistor Space}
\vskip 10pt

\baselineskip17truept
\centerline{Iosif Bena and Zvi Bern}
\baselineskip12truept
\centerline{\it Department of Physics and Astronomy}
\centerline{\it University of California, Los Angeles, Calif.\ 90095--1547}
\centerline{\tt iosif,bern@physics.ucla.edu}

\vskip 10pt
\centerline{David A. Kosower}
\baselineskip12truept
\centerline{\it Service de Physique Th\'eorique${}^{\natural}$,
            CEA--Saclay}
\centerline{\it F-91191 Gif-sur-Yvette cedex, France}
\centerline{\tt kosower@spht.saclay.cea.fr}

\vskip 10pt
\centerline{and}
\vskip 10pt

\centerline{Radu Roiban}
\baselineskip12truept
\centerline{\it Department of Physics}
\centerline{\it University of California, Santa Barbara, Calif.\ 93106}
\centerline{and}
\centerline{\it Department of Physics}
\centerline{\it Princeton University, Princeton, New Jersey 08544}
\centerline{\tt rroiban@princeton.edu}

\vskip 0.2in\baselineskip13truept

\vskip 0.5truein
\centerline{\bf Abstract}
{\narrower 

We elucidate the one-loop twistor-space structure corresponding to
momentum-space MHV diagrams. We also discuss the infrared divergences,
and argue that only a limited set of MHV diagrams contain them.  We
show how to introduce a twistor-space regulator corresponding to
dimensional regularization for the infrared-divergent diagrams.  We
also evaluate explicitly the `holomorphic anomaly' pointed out by
Cachazo, Svr\v cek, and Witten, and use the result to define modified
differential operators which can be used to probe the twistor-space
structure of one-loop amplitudes.

}
\vskip 0.3truein


\vfill
\vskip 0.1in
\noindent\hrule width 3.6in\hfil\break
\noindent
${}^{\natural}$Laboratory of the
{\it Direction des Sciences de la Mati\`ere\/}
of the {\it Commissariat \`a l'Energie Atomique\/} of France.\hfil\break

\Date{}

\line{}

\baselineskip17pt
%

\section{Introduction}
\vskip 10pt

\def\CP{{\mathbb CP}}
\def\C{{\mathbb C}}

In a recent paper~[\use\WittenTopologicalString], Witten suggested
that tree-level amplitudes of non-Abelian gauge theories can be
obtained by integrating over the moduli space of certain
$D$-instantons in the open-string topological B-model on (super)
twistor space $\CP^{3|4}$. This proposal generalizes Nair's earlier
construction~[\use\Nair] of maximally helicity-violating (MHV)
amplitudes~[\use\ParkeTaylor,\use\RecurrenceRelations].  It may help
shed light on the relative simplicity of amplitudes in unbroken gauge
theories.  Cachazo, Svr\v cek, and Witten (CSW)~[\use\CSW] distilled
the twistor-space structure underlying gauge-theory amplitudes in
Witten's proposal into a novel method to construct arbitrary
tree-level amplitudes, using off-shell continuations of MHV
amplitudes.  Although no-one has yet given a direct derivation of the
CSW construction from a Lagrangian, the combination of the correct
pole structure and Lorentz invariance leaves little doubt that it is
correct.  Related investigations and applications of the CSW
construction have appeared in
refs.~[\use\CSWApplications,\use\RecursiveTwistor,\use\NMHV].

The success of the twistor-space picture at tree level leads one to
investigate whether it can be extended to loop corrections.  In the
twistor-space string theory (or theories\footnote{${}^*$}{It is not
clear to us that the three theories discussed in
refs.~[\use\WittenTopologicalString,%
\use\Berkovits,\use\Vafa]
are in fact identical.}) proposed by
Witten~[\use\WittenTopologicalString], Berkovits~[\use\Berkovits], and
Neitzke and Vafa~[\use\Vafa], loop corrections appear to include
conformal supergravitons in an inextricable way.
(Siegel~[\use\Siegel] has proposed an alternative approach.)  However,
a more straightforward and direct approach to this question was
proposed by Brandhuber, Spence, and Travaglini~(BST)~[\use\BST].  They
were able to map one-loop diagrams built out of two MHV vertices to
the correct one-loop result for MHV amplitudes~[\use\NeqFourOneLoop],
by showing that their cuts map directly onto the cuts used in
calculating the amplitudes.  We will call such diagrams twistor-rules
diagrams.  (A more direct calculation of such loop diagrams without
reference to the cuts fails to give the correct answer, presumably
because of lack of a correct $i\varepsilon$ prescription.)  Abe, Nair,
and Park's construction~[\use\NairEffectiveTheory] of an effective
Lagrangian points in the same direction.

The BST calculation does not yet have a twistor-space derivation, but
it may be taken as `experimental' evidence that such a derivation
should exist.  This leads to the question: what is the twistor-space
picture corresponding to their calculation?  At tree level, the
twistor-space picture corresponding to the CSW construction takes sets
of points on various lines (each corresponding to one MHV vertex), and
connects these lines by twistor-space propagators in such a way as to
obtain a connected tree-level topology.  At tree level, one can
further argue~[\use\Gukov] that this construction is equivalent to a
picture in which all propagators collapse to points, forcing the
attached lines to intersect.  This confirms the original observation
of Cachazo, Svr\v cek, and Witten, based on known results for the
amplitudes, that a tree-level twistor-space amplitude is supported on
a locus of intersecting lines.  Gukov et al.~also argue that this
construction is then equivalent to one in which the points lie on a
higher-degree curve in twistor space.  (Three of the authors presented
a field-theory picture of this equivalence~[\use\RecursiveTwistor].)
This matches the calculations done by Spradlin, Volovich, and one of
the authors~[\use\RSV], and explains why the one-instanton calculation
is equivalent to a disconnected multi-instanton one.

The explicit calculations suggest that in twistor space, one-loop
amplitudes are supported on lines connected by propagators, the
ensemble forming a connected genus-one topology.  We will show that
the BST form matches this intuition.  This may not seem to be in
accord with the configurations found by Cachazo, Svr\v cek, and Witten in
their paper on MHV loop amplitudes~[\use\CSWLoop].  While they found
that the infrared-divergent contributions are supported precisely on
configurations of two lines connected by propagators, it appeared that
the finite parts are supported on configurations where one propagator
has collapsed, and one other point has wandered off its original line
into the plane defined by the two intersecting lines.  As the same
authors have argued more recently~[\use\CSWLoopII], however, there is
a subtlety in the use of momentum-space differential operators to
establish this picture.  We will compute explicitly the `holomorphic
anomaly' they point out, and show that when it is taken into account,
the twistor-space picture of the amplitudes is indeed the one
suggested by the twistor-space partner to the BST calculation as
presented herein.  One can define modified differential operators,
subtracting the anomaly contribution, which do annihilate the one-loop
amplitude.

In the next section we review in detail the relation between a tree-level
momentum-space amplitude and the corresponding twistor-space amplitude.
Before following a similar path for one-loop amplitudes, we must understand
where their infrared divergences emerge; we do so in 
sect.~\use\InfraredDivergenceSection.  We then give the twistor-space
counterpart of infrared-finite twistor-rules diagrams in 
sect.~\use\InfraredFiniteSection, and the infrared-divergent ones in
sect.~\use\InfraredDivergentSection.  We also show how to regulate the
divergences in twistor space in a manner compatible with ordinary
dimensional regularization in momentum space.  
In sect.~\use\DifferentialOperatorSection, we compute the 
`holomorphic anomaly' explicitly, and show that one can use it
to define a modified differential operator for probing
the twistor-space structure of amplitudes.  In 
appendix~\use\FourierAppendix, we discuss a delta-convergent
sequence useful for dimensional regularization in twistor space; and
in appendix~\use\HolomorphicAppendix, we discuss details of the
holomorphic anomaly computation.

\section{Tree-Level Amplitudes}
\tagsection\TreeLevelAmplitudeSection
\vskip 10pt

\topinsert\LoadFigure\NMHVDiagramFigure
{\baselineskip 13 pt
\noindent\narrower
An example of a CSW diagram contributing to the tree amplitude
with three negative-helicity gluons.
}  {\epsfxsize 2.7 truein}{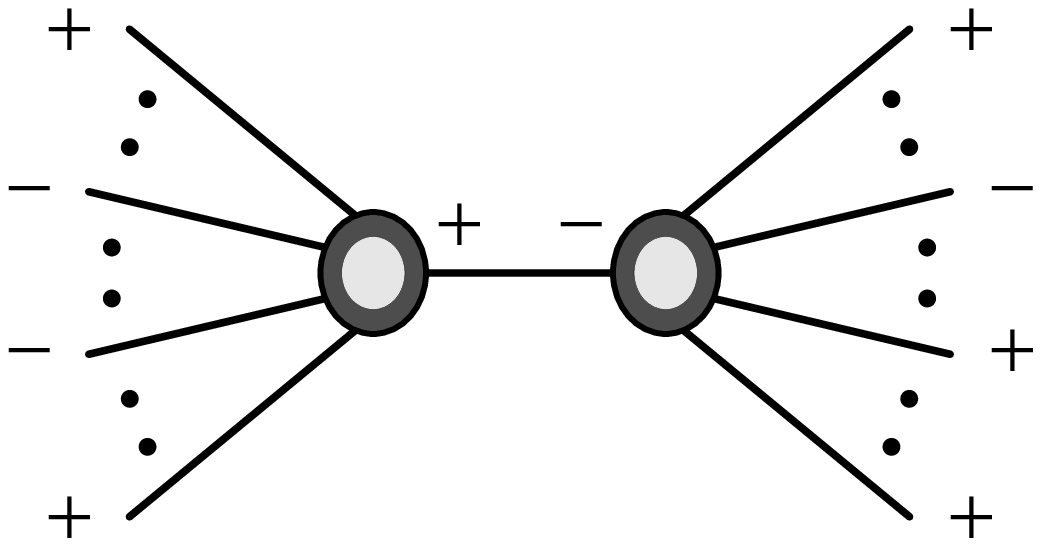}{}\endinsert

We begin by reviewing the connection between tree-level amplitudes in
momentum space, and the twistor-space amplitudes emerging from
Witten's string theory.  Rather than starting from a twistor-space
amplitude, we begin from a momentum-space amplitude and work backwards
to the twistor-space one.  As our example, we take a diagram $\delta\!
A$ contributing to an amplitude with three negative-helicity gluons or a
next-to-MHV (NMHV) amplitude. This is the simplest amplitude
containing more than one MHV vertex. This contribution is, 
$$\eqalign{
(2\pi)^4\delta^4(P_A+P_B)\delta\! A(P_A,P_B) &= 
(2\pi)^4\delta^4(P_A+P_B)  V(P_A, -P_A^\proj) {i\over P_A^2} V(P_B,-P_B^\proj)
\cr &=
(2\pi)^4\int d^4k_1 d^4k_2\;\delta^4(P_A+k_1)
   V(P_A, k_1^\proj)\frac{i\delta^4(k_1+k_2)}{k_1^2+i\varepsilon} 
\cr&\hphantom{=!} \hskip 20mm\times
     \delta^4(P_B+k_2)V(P_B, k_2^\proj),
}\eqn\CSWTreeRep$$
corresponding to the diagram shown in fig.~\use\NMHVDiagramFigure.  In this
equation, $V(P,k)$ is an MHV vertex depending on the momenta in the set
$P$, and on the off-shell momentum $k$.  The off-shell continuation
of the MHV amplitudes to MHV vertices is defined using the projected
momentum~[\use\NMHV,\use\RecursiveTwistor],
\def\adot{{\dot a}}
\def\bdot{{\dot b}}
$$
k_{a\adot}^\proj = k_{a\adot}-{k^2\over 2 k\cdot \eta} \eta_{a\adot},
\anoneqn$$
where $\eta$ can be thought of as the light-cone vector of light-cone
gauge.  (This is equivalent to the CSW off-shell continuation.)
This motivates the BST decomposition,
\def\tl{\smash{\tilde l}}
\def\teta{\tilde\eta}
\def\tlambda{\tilde\lambda}
$$
k_{a\adot} = l_a \tl_\adot + z \eta_a \teta_\adot.
\eqn\BSTDecomposition$$
Each momentum $p_{Xi} \in P_{X}$, can be decomposed in terms of spinorial
variables,
$$
p_{Xi\,a\adot} = \lambda_{Xi\,a}\tlambda_{Xi\,\adot}.
\anoneqn$$
The vertices can then be thought of as functions of the $\lambda$ and
$\tlambda$.  The simplification captured by the CSW rules arises because
the vertices are in fact independent of the $\tlambda_i$,
$$
V = V(\lambda,l).
\anoneqn$$
The $i\varepsilon$ in the propagator does not play a role at tree level,
but will at loop level.

\def\twist{{\rm T}}
The twistor-space amplitude $A^{\twist}$ is given by 
a `half-Fourier transform' of the momentum-space one, that is a Fourier
transform with respect to the $\tlambda$ variables alone,
$$
A^\twist(\lambda_A,\lambda_B,\mu_A,\mu_B) =
(2\pi)^4\int\prod_{i\in A}{d^2{\tlambda}_{Ai\,\adot}\over (2\pi)^2}\;
e^{i[\mu_{Ai},{\tlambda}_{Ai}]}
\prod_{i\in B}{d^2{\tlambda}_{Bi\,\bdot}\over (2\pi)^2}\;
e^{i[\mu_{Bi},{\tlambda}_{Bi}]}\delta^4(P_A+P_B)\,A(P_A,P_B).
\anoneqn$$
(Note that the sign conventions on the spinor products used here and
elsewhere in the twistor-space literature are different from
those used conventionally in the QCD literature.  We will use the notation
$\tspa{\lambda_1}.{\lambda_2}$ and 
$\tspb{\smash\tlambda_1}.{\smash\tlambda_2}$
in the present paper in contrast to the comma-less (and $\lambda$-less)
notation $\spa1.2$ and $\spb1.2$ standard in the QCD literature.  While
the two angle products agree, 
$\tspb{\smash\tlambda_1}.{\smash\tlambda_2} = -\spb{1}.{2}$.)

Using the CSW representation~(\use\CSWTreeRep), and 
rewriting the delta functions involving the external momenta, we obtain
for the diagram of eq.~(\use\CSWTreeRep),
$$\eqalign{
&(2\pi)^4\int \frac{d^4k_1}{(2\pi)^4} \frac{d^4k_2}{(2\pi)^4} 
       \frac{i\delta^4(k_1+k_2)}{k_1^2+i\varepsilon}
  \int d^4x_A^{a\adot} d^4x_B^{b\bdot}
\int\prod_{i\in A}{d^2{\tlambda}_{Ai\,\adot}\over(2\pi)^2}
\prod_{i\in B}{d^2{\tlambda}_{Bi\,\bdot}\over(2\pi)^2}\;
e^{i[\mu_{Ai},{\tlambda}_{Ai}]+i[\mu_{Bi},{\tlambda}_{Bi}]}
\cr &\hphantom{ d^4k_1 }\times
\exp\Bigl[ix_A^{a\adot}
  \bigl(\sum_{i\in A}\lambda_{Ai\,a}\tlambda_{Ai\,\adot}
                     +l_{1\,a}\tl_{1\,\adot}
                     +z_1\eta_a\teta_{\adot}\bigr)\Bigr]\,V(\lambda_A,l_1)
\cr &\hphantom{ d^4k_1 }\times 
\exp\Bigl[ix_B^{b\bdot}
  \bigl(\sum_{i\in B}\lambda_{Bi\,b}\tlambda_{Bi\,\smash{\bdot}}
                     +l_{1\,b}\tl_{1\,\smash{\bdot}}
                 +z_2\eta_b\teta_{\smash{\bdot}}\bigr)\Bigr]\,V(\lambda_B,l_2).
}\anoneqn$$
Performing the $\tlambda_i$ integrals yields 
delta functions for the $\mu_{Xi}$,
$$\eqalign{
&(2\pi)^4 \int  \frac{d^4k_1}{(2\pi)^4} \frac{d^4k_2}{(2\pi)^4} 
       \frac{i\delta^4(k_1+k_2)}{k_1^2+i\varepsilon}
  \int d^4x_A^{a\adot} d^4x_B^{b\bdot}
\int d^2 m_1^{\adot} d^2m_2^{\smash{\bdot}}\,
e^{i[m_1,\tl_1]+i[m_2,\tl_2]}
\cr &\hphantom{ d^4k_1 }\times
\exp\Bigl[ix_A^{a\adot} z_1\eta_a\teta_{\adot}\Bigr]\,V(\lambda_A,l_1)
   \prod_{i\in A} \delta^2(\mu_{Ai}^{\adot} - x_A^{a\adot}\lambda_{Ai\,a})
   \delta^2(m_{1}^{\adot} - x_A^{a\adot}l_{1\,a})
\cr &\hphantom{ d^4k_1 }\times 
\exp\Bigl[ix_B^{b\bdot} z_2\eta_b\teta_{\smash{\bdot}}\Bigr]\,
      V(\lambda_B,l_2),
   \prod_{i\in B} \delta^2(\mu_{Bi}^{\smash{\bdot}} 
                           - x_B^{b\smash{\bdot}}\lambda_{Bi\,b})
   \delta^2(m_{2}^{\adot} - x_A^{a\adot}l_{2\,a}),
}\anoneqn$$
where we have also introduced new variables $m_{1,2}$ conjugate
to $\tl_{1,2}$.
Performing the $k_2$ integral will set $l_2\tl_2 = -l_1\tl_1$ and
$z_2 = -z_1$; making the phase choice $l_2 = l_1$ 
(and hence $\tl_2 = -\tl_1$), we obtain the expression,
$$\eqalign{
&\int \frac{d^4k}{(2\pi)^4} {i\over k^2+i\varepsilon}
  \int d^4x_A^{a\adot} d^4x_B^{b\bdot}\;
\int d^2 m_1^{\adot} d^2m_2^{\smash{\bdot}}\,
e^{i[(m_1-m_2),\tl]}
\cr &\hphantom{ d^4k_1 }\times
\exp\Bigl[i(x_A-x_B)^{a\adot} z\eta_a\teta_{\adot}\Bigr]\,V(\lambda_A,l)
      V(\lambda_B,l)
\cr &\hphantom{ d^4k_1 }\times 
   \prod_{i\in A} \delta^2(\mu_{Ai}^{\adot} - x_A^{a\adot}\lambda_{Ai\,a})
   \delta^2(m_{1}^{\adot} - x_A^{a\adot}l_{a})
   \prod_{i\in B} \delta^2(\mu_{Bi}^{\smash{\bdot}} 
                           - x_B^{b\smash{\bdot}}\lambda_{Bi\,b})
   \delta^2(m_{2}^{\adot} - x_A^{a\adot}l_{a}),
}\eqn\TwistorAmplitudeA$$
for the twistor-space amplitude (we have dropped the subscript `1' on
$k$, $l$, and $\tl$).

Our final task is to convert the momentum space propagator to a twistor-space
one.  We make use of the decomposition~[\use\BST] of the loop-integral
measure
 into the Nair measure~[\use\Nair] and a dispersive measure over $z$,
$$
{d^4k\over k^2 + i\varepsilon} = {dz\over z+i\varepsilon}
 \Bigl[ \tspa{l}.{dl} d^2\tl - \tspb{l}.{dl} d^2l\Bigr].
\anoneqn$$
In the twistor-space form, the integral over $z$ is just a theta function,
$$
I_\eta(x_A-x_B) = 
   -i\int_{-\infty}^{\infty} {dz\over 2\pi} {1\over z+i\varepsilon}\;
\exp\Bigl[i(x_A-x_B)^{a\adot} z\eta_a\teta_{\adot}\Bigr],
\anoneqn$$
which restricts the integral to half of the moduli space,
while the $l$ and $\tl$ integrals,
$$
G_0(m_1,m_2;l) = -\int {1\over(2\pi)^3}
  \Bigl[ \tspa{l}.{dl} d^2\tl - \tspb{l}.{dl} d^2l\Bigr]
\exp\Bigl[{i[(m_1-m_2),\tl]}\Bigr],
\eqn\ProtoPropagator$$ should yield the twistor-space 
propagator.  To see this we would like to transform the measure to
the CSW form.  We can do this as follows.  Define a real scale factor $\tau$,
$$
l = \tau l', \qquad
\tl = \tau \tl',
\anoneqn$$
\def\wb{{\overline w}}
where
$$
l' = (1,w),\qquad
\tl' = (1,\wb)
\anoneqn$$
($\wb$ will be taken to be the complex conjugate of $w$).
Then 
$$\eqalign{
&\tspa{l}.{dl} = \tau^2 \tspa{l'}.{dl'}, \qquad
\tspb{\tl}.{d\tl} = \tau^2 \tspb{\tl'}.{d\tl'},
\cr & 2 d^2\tl = \tspb{d\tl}.{d\tl} = 2 \tau d\tau \tspb{\tl'}.{d\tl'},\qquad
2 d^2l = \tspa{dl}.{dl} = 2\tau d\tau \tspa{l'}.{dl'},
}\anoneqn$$
and
$$  \Bigl[ \tspa{l}.{dl} d^2\tl - \tspb{l}.{dl} d^2l\Bigr]
=2\tau^3 d\tau \wedge \tspa{l'}.{dl'} \wedge \tspb{\tl'}.{d\tl'},
\eqn\CSWMeaure$$
which is the CSW measure ($t_{\rm CSW} = \tau^2$).

Now, the vertices $V$ are homogeneous functions of the spinors
$\lambda_{Xi\,a}$.  Also, the two opposite ends of the propagator
line correspond to gluons of opposite helicity exiting the two vertices,
and so,
$$
V(\lambda_A,l) V(\lambda_B,l) = 
V(\lambda_A,l') V(\lambda_B,l').
\anoneqn$$
We can therefore ignore the vertices in evaluating $\tau$ and $\tl'$
integrals.  Write out the protopropagator~(\use\ProtoPropagator)
 in terms of $\tau$ and
$\wb$, 
$$
G_0 = -{1\over (2\pi)^3}\int_{-\infty}^\infty \tau^3 d\tau\;\int_{\C} dw d\wb\;
\exp\Bigl[i \tau^2 [m_{12}^{\prime\smash{\dot1}} \wb
          -m_{12}^{\prime\smash{\dot2}}]
          \Bigr],
\eqn\PropagatorA$$
where we have also used rescaled $m_i' = m_i/\tau$ and denoted
$m_1'-m_2'$ by $m_{12}'$.
Factors of $\tau$ from the rescaling cancel between the $d^2m$ measure
and the delta functions requiring that the endpoints of the propagator
lie on the $x_{A,B}$ lines.
We can evaluate the $\wb$ integral, which yields a delta function,
$$\eqalign{
2\pi\delta\bigl(\tau^2 m_{12}^{\prime\smash{\dot1}}\bigr) =
{2\pi\over \tau^2}\delta\bigl((m_1'-m_2')^{\smash{\dot1}}\bigr),
}\anoneqn$$
where the equality follows from the factors of $\tau$ in the integration
measure in eq.~(\use\PropagatorA).  In order to do the $\tau$ integral,
we must introduce an $i\varepsilon$ prescription for 
$m_{12}^{\prime\smash{\dot2}}\rightarrow 
m_{12}^{\prime\smash{\dot2}}-i\varepsilon$.
If we now do the $\tau$ integral,
we obtain
$$\eqalign{
 & {i\over2\pi}{\delta(m_{12}^{\prime\smash{\dot1}})
        \over m_{12}^{\prime{\dot2}}-i\varepsilon} 
     \int_{-\infty}^{\infty} {dw\over 2\pi}
\cr &= {i\over2\pi}{\delta(m_{12}^{\prime\smash{\dot1}})
        \over m_{12}^{\prime{\dot2}}-i\varepsilon} 
                \int_{-\infty}^{\infty} {dw_1\over 2\pi} {dw_2\over 2\pi}\;
                             2\pi\delta(w_1-w_2).
}\eqn\Propagator$$
We cannot do the $w$ integral, because the vertices do depend
on this variable.  
Indeed, these remaining integrals represent the $l$ integrations
over the endpoints of the propagator, and should be thought of as part 
of the moduli space of curves for this amplitude.
The delta function $\delta(w_1-w_2)$ on the second line should be considered
 part of the propagator. 

\topinsert\LoadFigure\TwistorTreeFigure
{\baselineskip 13 pt
\noindent\narrower
The twistor-space configuration contributing to an NMHV amplitude.
The external points lie on a pair of lines connected by a twistor-space
propagator, represented by a dashed line.
}  {\epsfxsize 1.2 truein}{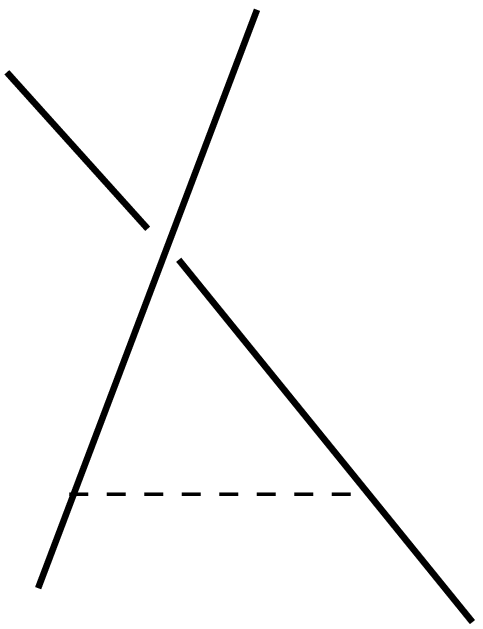}{}\endinsert
Define a propagator without the $w$ integral,
$$ G(m_{12}') \equiv 
{i\over2 \pi}{\delta(m_{12}^{\prime\smash{\dot1}})
        \over m_{12}^{\prime\smash{\dot2}}-i\varepsilon} .
\eqn\PropagatorI$$
Alternatively, averaging over both order of evaluation of 
the $\tau$ and ${\bar w}$ integrals in eq.~(\PropagatorA), we obtain
$$
\eqalign{
G_2(m_{12}')={i\over 4\pi}
\left[
{\delta(m_{12}^{\prime\smash{\dot1}})
        \over m_{12}^{\prime{\dot2}}-i\varepsilon} 
+
{\delta(m_{12}^{\prime\smash{\dot2}})
        \over m_{12}^{\prime{\dot1}}-i\varepsilon} 
\right]~~.
}\eqn\PropagatorPV
$$
(In the second order of evaluation we have used a principal-value 
prescription.) At tree level both eq.~(\PropagatorI) and  eq.~(\PropagatorPV)
lead to the same answer.
Inserting the former into eq.~(\use\TwistorAmplitudeA), we obtain
the final expression for the tree-level diagram,
$$\eqalign{
& \int d^4x_A^{a\adot} d^4x_B^{b\bdot}
\int {dl'\over 2\pi} d^2 m_1^{\prime\adot} d^2m_2^{\prime\smash{\bdot}}\,
\cr &\hphantom{ d^4k_1 }\times
V(\lambda_A,l') V(\lambda_B,l') I_\eta(x_A-x_B) G(m_{12}')
\cr &\hphantom{ d^4k_1 }\times 
   \prod_{i\in A} \delta^2(\mu_{Ai}^{\adot} - x_A^{a\adot}\lambda_{Ai\,a})
   \delta^2(m_{1}^{\prime\adot} - x_A^{a\adot}l'_{a})
   \prod_{i\in B} \delta^2(\mu_{Bi}^{\smash{\bdot}} 
                           - x_B^{b\smash{\bdot}}\lambda_{Bi\,b})
   \delta^2(m_{2}^{\prime\adot} - x_A^{a\adot}l'_{a}).
}\anoneqn$$
This result is equivalent to the integrals considered in
sect.~6 of ref.~[\use\CSW]. 
It expresses this diagram's contribution to the twistor-space amplitude
as a function supported on a pair of lines connected by a propagator,
illustrated in fig.~\use\TwistorTreeFigure.

\section{Hunting Infrared Divergences}
\tagsection\InfraredDivergenceSection
\vskip 10pt

From the results of ref.~[\use\BST] we may expect, as also noted in
ref.~[\use\CSWLoopII], that the simplest loop amplitude in twistor space 
(an MHV amplitude) would be obtained by connecting the two lines 
in the tree-level amplitude of sect.~\use\TreeLevelAmplitudeSection{} by
another propagator, so as to obtain a genus-one configuration.  This
will indeed turn out to be the case.  To reach such a representation,
however, we must first confront the infrared singularities in the amplitude.

These infrared singularities are conventionally regulated dimensionally
in perturbative gauge theories.  In momentum space,
they arise from configurations where
the loop momentum is nearly on shell and either soft or collinear with
an external momentum (or both).  When computing in $D=4-2\e$ dimensions,
with $\e<0$, the singularities manifest themselves in the appearance of
poles in $\e$.  In gauge theories, the leading singularities are
a factor of $1/\e^2$ per loop order.

The one-loop amplitude for the MHV amplitude in ${\cal N}=4$ supersymmetric
gauge theory was computed a decade ago, by Dixon, Dunbar, and two of the
authors~[\use\NeqFourOneLoop].  The leading-color or planar 
contributions\footnote{${}^*$}{The subleading-color terms can be obtained
algebraically from the leading-color ones as discussed in sect.~7 of
ref.~[\use\NeqFourOneLoop].}
have the form of a sum over color permutations, with each term in the sum
equal to,
\def\easytwo{{{\rm 2m}\,e}}
\def\onemass{{\rm 1m}}
\def\cg{c_\Gamma}
$$
A^\oneloop_n = \cg A_n^\tree\, V_n^g,
\anoneqn$$
where
$$\eqalign{
\cg &= {1\over (4\pi)^{2-\e}}{\Gamma(1+\e)\Gamma^2(1-\e)\over\Gamma(1-2\e)}
\cr
V^g_{2m+1} &= (\mu^2)^{\e} \sum_{r=2}^{m-1}\sum_{i=1}^n 
F^\easytwo_{n:r;i} + \sum_{i=1}^n F^\onemass_{n:i},
\cr V^g_{2m} &= (\mu^2)^{\e} \sum_{r=2}^{m-2}\sum_{i=1}^n 
F^\easytwo_{n:r;i} + \sum_{i=1}^n F^\onemass_{n:i}
+ \sum_{i=1}^{n/2} F^\easytwo_{n:m-1;i}.
}\anoneqn$$

The integral functions $F^\easytwo$ and $F^\onemass$ are given in terms of 
$$\eqalign{
F(s,t,P^2,Q^2) &= -{1\over\e^2} \biggl[
 \bigl({ -s}\bigr)^{-\e}
 +\bigl({ -t}\bigr)^{-\e}
 -\bigl({ -P^2}\bigr)^{-\e}
 -\bigl({ -Q^2}\bigr)^{-\e}
\biggr]
\cr &\hphantom{=!} 
+\Li_2\Bigl(1-{P^2\over s}\Bigr)
+\Li_2\Bigl(1-{P^2\over t}\Bigr)
+\Li_2\Bigl(1-{Q^2\over s}\Bigr)
+\Li_2\Bigl(1-{Q^2\over t}\Bigr)
\cr &\hphantom{=!} 
-\Li_2\Bigl(1-{P^2 Q^2\over s t}\Bigr)
+{1\over2} \ln^2 \Bigl({s\over t}\Bigr),
}\anoneqn$$
as follows,
$$\eqalign{
F^\onemass_{n:i} &= F(s_{(i-3)(i-2)},s_{(i-2)(i-1)},s_{i\cdots(i+n-4)},0),\cr
F^\easytwo_{n:r;i} &= F(s_{(i-1)\cdots(i+r-1)},s_{i\cdots(i+r)},
                        s_{i\cdots (i+r-1)},s_{(i+r+1)\cdots(i-2)}),\cr
}\anoneqn$$
where $s_{j_1\cdots j_2} = K_{j_1\cdots j_2}^2 = (k_{j_1}+\cdots+k_{j_2})^2$
and all indices are understood cyclicly $\mod n$.
($F$ has a smooth limit as $P^2$ or $Q^2\rightarrow 0$.)
These formul\ae{} 
make it appear as though {\it all\/} diagrams will contribute
infrared-divergent terms.

Appearances can be deceiving, however, and in this case, they are.
Each diagram built out of MHV vertices corresponds
closely to a cut in a specific channel in
a unitarity-based calculation~[\use\NeqOneOneLoop,\use\UnitarityMachinery], 
which uses on-shell amplitudes
on both sides of a cut.  This close correspondence allows us to show
that {\it most\/} MHV diagrams are in fact free of infrared divergences,
which are present only in a subset of diagrams.  Indeed, if we examine
a generic cut of the amplitude, say in the $s_{c_1\cdots c_2}$ channel
($|c_1-c_2|>1$),
we find,
$$\eqalign{
&2\pi i \cg A^\tree \cr
&\hskip 2mm\times\ln\biggl({ (s_{(c_1-1)\cdots (c_2-1)} s_{c_1\cdots c_2}
                     - s_{c_1\cdots (c_2-1)} s_{(c_1-1)\cdots c_2})
                    (s_{c_1\cdots c_2} s_{(c_1+1)\cdots(c_2+1)}
                     - s_{c_1\cdots (c_2+1)} s_{(c_1+1)\cdots c_2})
                   \over 
                    (s_{c_1\cdots (c_2-1)} s_{(c_1+1)\cdots c_2}
                     -s_{c_1\cdots c_2} s_{(c_1+1)\cdots (c_2-1)})
                    (s_{(c_1-1)\cdots c_2} s_{c_1\cdots (c_2+1)}
                     -s_{c_1\cdots c_2} s_{(c_1-1)\cdots (c_2+1)})
                  }\biggr),
}\eqn\InfraredFiniteCut$$
a finite result.  But the infrared-divergent contributions to the
amplitude arise from
infrared-divergent contributions to the cuts --- one power of $1/\e$
arises from the cut integral, and the other from the dispersion integral.
As shown by Brandhuber, Spence, and Travaglini, however, the cut of the
twistor-rules diagram built out of the two MHV vertices 
$V(\ell_1,c_1,\ldots,c_2,\ell_2)$ and
$V(-\ell_2,c_2+1,\ldots,c_1-1,-\ell_1)$ is exactly the same as the above
cut.  This in turn implies that the twistor-rules diagram does not
have infrared divergences either.  

Indeed, we know on general grounds~[\use\KSTir] 
that the infrared divergences of
a one-loop diagram in the ${\cal N}=4$ theory are of the form,
$$
V_{n,{\rm IR}}^g = -{(\mu^2)^\e\over\e^2} \sum_{i=1}^n (-s_{i(i+1)})^{-\e},
\anoneqn$$
involving only nearest-neighbor two-particle invariants.  In the
MHV amplitude, these arise
from twistor-rules diagrams where one of the vertices is a four-point
vertex.

We can understand this result as well~[\use\DAKir] by looking at the
cuts of diagrams.  Infrared divergences in the cuts can arise only
from regions where the phase-space integral diverges as $\e\rightarrow 0$.
For generic values of the external momenta, singularities
in multiparticle invariants are at isolated points, 
and hence of insufficient dimensionality to induce such 
a divergence.  Only two-particle invariants will produce such
a divergence.  However, 
in general the only two-particle singularities involving the cut loop
momentum are collinear singularities, where that momentum becomes
collinear to a neighboring external momentum.  These 
are characteristically $1/\sqrt{\vphantom{A}s_{ij}}$ singularities, which are 
integrable even as $\e\rightarrow 0$.  (Here, it is crucial that
each side of the cut is an {\it amplitude\/}, and not merely a
Feynman tree diagram; in most gauges,
this statement is {\it not\/} true diagram
by diagram.)  

In the particular case of the MHV loop amplitude,
in the channel we are examining, there are four candidate regions,
corresponding to the inverse spinor products
$$
{1\over\tspa{\ell_1}.{c_1}},
\quad {1\over\tspa{\ell_1}.{(c_1-1)}},
\quad {1\over\tspa{\ell_2}.{(c_2+1)}},
\quad {\rm and\ } {1\over\tspa{\ell_2}.{c_2}},
\eqn\SingularitiesI$$
where $\ell_i$ are the cut momenta.  Each of these expressions is singular
in collinear regions ($\ell_1\parallel k_{c_1}$, etc.), 
but that singularity only goes as 
$1/\theta_{\ell c_1}$.  The measure 
is $d(\cos\theta) \sim \theta d\theta$, so the expressions
are integrable, and give rise to no infrared singularity. 

The exception is of course the cut where the amplitude on one side
is a four-point amplitude.  In this case,
when one of the cut loop momenta becomes collinear to the neighboring
external momentum, momentum conservation forces the other cut loop momentum
to become collinear to {\it its\/} neighboring external momentum, and
both $1/\sqrt{\vphantom{A} s_{ij}}$ singularities overlap to 
produce a $1/s_{ij}$-type
singularity or a $d\theta/\theta$ integral, 
which does give rise to a $1/\e$ pole when regulated dimensionally.
  (This singularity in the tree amplitude
should be characterized as a forward-scattering singularity rather than
as a collinear singularity.)

The twistor-rules diagrams contributing to the one-loop amplitude thus
fall into two classes: (a) diagrams containing a four-point vertex, which
are infrared divergent, and require an implementation of dimensional
regularization, and (b) all other diagrams, which are infrared-finite,
and hence can be treated in four dimensions.  (Note that diagrams with
a three-point vertex have vanishing cut and so can be ignored.)
We consider the twistor-space
form of the latter diagrams in the next section.

\section{Infrared-Finite Contributions to One-Loop Amplitudes}
\tagsection\InfraredFiniteSection
\vskip 10pt

We can obtain the expression for a single one-loop twistor-rules diagram 
from the tree-level one~(\use\CSWTreeRep)
by adding one additional propagator, along with
appropriate delta functions to produce a genus-one configuration,
$$\eqalign{
\int d^4k_{A1} d^4k_{B1} &d^4k_{A2} d^4k_{B2}\;\delta^4(k_{A1}+P_A+k_{A2})
   V(k_{A1}^\proj,P_A, k_{A2}^\proj)
       \frac{i\delta^4(k_{A1}+k_{B1})}{k_{A1}^2+i\varepsilon} 
       \frac{i\delta^4(k_{A2}+k_{B2})}{k_{A2}^2+i\varepsilon} 
\cr&\times
     \delta^4(k_{B1}+P_B+k_{B2})V(k_{B2}^\proj,P_B, k_{B1}^\proj).
}\eqn\CSWOneLoopRep$$
In this section, we restrict attention to diagrams where both vertices
have more than four legs (including the sewn legs).  As we have seen
above, these integrals are infrared-convergent and can be treated in
four dimensions.  

\topinsert\LoadFigure\TwistorMHVLoopFigure
{\baselineskip 13 pt
\noindent\narrower
The twistor-space configuration contributing to an MHV amplitude.
The external points lie on a pair of lines connected by two twistor-space
propagators.
}  {\epsfxsize 1.2 truein}{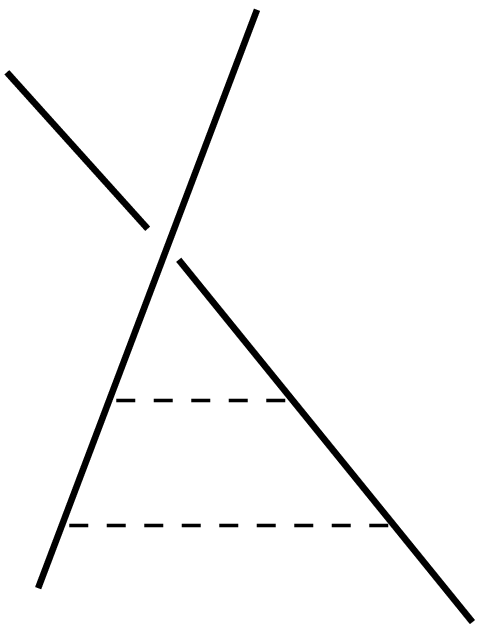}{}\endinsert

Following similar manipulations as in sect.~\use\TreeLevelAmplitudeSection,
we find the following expression for the contribution of
a twistor-rules diagram to the one-loop
twistor-space amplitude,
$$\eqalign{
& \int d^4x_A^{a\adot} d^4x_B^{b\bdot}\;
\int {dl_{1}'\over 2\pi} {dl_{2}'\over 2\pi}\,
\int d^2 m_{A1}^{\prime\adot} d^2m_{A2}^{\prime\adot}\,
\int d^2 m_{B1}^{\prime\smash{\bdot}} d^2m_{B2}^{\prime\smash{\bdot}}\,
\cr &\hphantom{ d^4k_1 }\times
V(l_{1}',\lambda_A,l_{2}') V(l_{2}',\lambda_B,l_{1}') I_\eta(x_A-x_B) 
G_2(m_{A1,B1}')
G_2(m_{A2,B2}')
\cr &\hphantom{ d^4k_1 }\times 
   \prod_{i\in A} \delta^2(\mu_{Ai}^{\adot} - x_A^{a\adot}\lambda_{Ai\,a})
   \delta^2(m_{A1}^{\prime\adot} - x_A^{a\adot}l'_{1\,a})
   \delta^2(m_{B1}^{\prime\bdot} - x_B^{b\bdot}l'_{1\,b})
\cr &\hphantom{ d^4k_1 }\times 
   \prod_{i\in B} \delta^2(\mu_{Bi}^{\smash{\bdot}} 
                           - x_B^{b\smash{\bdot}}\lambda_{Bi\,b})
   \delta^2(m_{A2}^{\prime\adot} - x_A^{a\adot}l'_{2\,a})
   \delta^2(m_{B2}^{\prime\bdot} - x_B^{b\bdot}l'_{2\,b}).
}\eqn\LoopEqn
$$
This expression gives the complete contribution for diagrams where
the two external gluons with negative helicity are attached to 
the same vertex, because only gluons can circulate in the loop. The use of $G_2$
reflects subtleties with the $i\epsilon$ prescription. For
other diagrams, we must include the contributions of the full
${\cal N}=4$ multiplet by sprinkling fermionic integrals and delta
functions into this equation.
This result expresses the contribution as a function supported on a pair
of lines connected by two propagators into a genus-one topology, illustrated
in fig.~\use\TwistorMHVLoopFigure.  This is a leading-color contribution,
as dictated by the ordering of points inside the two vertex functions $V$.

\section{Infrared-Divergent One-Loop Diagrams}
\tagsection\InfraredDivergentSection
\vskip 10pt

When the twistor-rules diagram does contain a four-point vertex,
it will be infrared divergent.  As we have discussed in 
sect.~\use\InfraredDivergenceSection, these divergences arise from
regions of the $l_i$ integration where $l_1$ becomes collinear to $\lambda_1$.
Momentum conservation then forces $l_2$ to become collinear to $\lambda_2$.
While it is not clear how to interpret the usual momentum-space regulator
($d^4 p\rightarrow d^D p$) in twistor space, it isn't necessary to have
such an interpretation.

One way to regulate the divergence is to smear out the energy-momentum
conserving delta function in eq.~(\use\CSWOneLoopRep), replacing
$$
\delta^4(k_{A1}+P_A+k_{A2}) \rightarrow \Delta^4_\xi(k_{A1}+P_A+k_{A2}),
\anoneqn$$
where $\Delta_\xi$ is a delta-convergent sequence, for example
$\Delta_\xi(x) = \xi/(\xi^2 + x^2)$.  In this case, one could perform
all the conversions to twistor space, and study the limit later.  The
infrared divergences manifest themselves as logarithmic divergences
in the $\xi\rightarrow 0$ limit.

We will take a different approach, and simply write down a regulator
in twistor space that is designed to reproduce the usual answer from
momentum-space dimensional regularization.  Suppose that it is set $A$
that is on the four-point vertex, and thus contains two external legs;
if these two are labeled $1$ and~$2$, then the diagram will result in
a contribution
$$
\sim (-s_{12})^{-\e} f(s_{12},\cdots;\e),
\anoneqn$$
where the leading behavior of $f$ is $-1/\e^2$.  We know that we need
to regulate the behavior of the $l_1$ integral near the collinear
region, softening the pole there\footnote{${}^\dagger$}{ Because we
have smeared out the energy-momentum conserving delta function by
representing it as an exponential integral over $x$, the $l_{1,2}$
integrals are not actually divergent.  However, if they are left
unregulated, the $x$ integral will turn out to be divergent.  Although
it may be possible to regulate the $x$ integral instead, it seems more
straightforward to regulate the $l_i$ integrals.}.  We can do so by
introducing a factor of $\tspa{l_1}.{\lambda_1}^{-\e}$, with $\e<0$,
into the integrand.  For symmetry, we also introduce a similar factor
for $l_2$.  If we then make the resulting factor homogeneous in the
$l_i$, we find,
$$
R_\e(l_{1},l_{2},\lambda_1,\lambda_2) =
\Bigl({\tspa{l_{1}}.{\lambda_1}\tspa{l_{2}}.{\lambda_2}\over
       \tspa{l_{1}}.{l_{2}}}\Bigr)^{-\e} ,
\anoneqn$$
for our regulating factor.  

We are not quite done.  The above factor will result in a pole contribution
proportional to $\tspa{\lambda_{1}}.{\lambda_{2}}^{-\e}$, 
which is not exactly
of the desired form; we are missing a factor of 
$\tspb{\smash\tlambda_1}.{\smash\tlambda_2}^{-\e}$.  
We can determine the required factor in the twistor-space integral
by multiplying the original momentum-space amplitude by this factor, 
and then performing the Fourier transforms with respect to
the $\tlambda$ variables as in sect.~\use\TreeLevelAmplitudeSection.
The computation of the Fourier transform is given in 
appendix~\use\FourierAppendix.
The result there shows that in order to introduce the desired factor
into the result, we should modify
the integrand, replacing the delta functions containing $m_{A1}$ and
$m_{A2}$ by a delta-convergent sequence,
$$
\delta^2(m_{A1}^{\prime\adot}-x_A^{a\adot}l'_a)
\delta^2(m_{A2}^{\prime\adot}-x_A^{a\adot}l'_a)
\rightarrow {\e\Gamma(2-\e)\over (2\pi)^3\Gamma(1+\e)} 
\tspb{\smash{m_{A1}'-x_A^{a}l'_{1\,a}}}.{\smash{m_{A2}'-x_A^{a}l'_{2\,a}}}^{\e-2}.
\anoneqn$$

The resulting expression for this loop diagram in twistor space is thus,
$$\eqalign{
& {\e\Gamma(2-\e)\over(2\pi)^3\Gamma(1+\e)}
   \int d^4x_A^{a\adot} d^4x_B^{b\bdot}\;
\int {dl_{1}'\over 2\pi} {dl_{2}'\over 2\pi}\,
\int d^2 m_{A1}^{\prime\adot} d^2m_{A2}^{\prime\adot}\,
\int d^2 m_{B1}^{\prime\smash{\bdot}} d^2m_{B2}^{\prime\smash{\bdot}}\,
\cr &\hphantom{ d^4k_1 }\times
R_\e(l_{1},l_{2},\lambda_1,\lambda_2)
V(l_{1}',\lambda_A,l_{2}') V(l_{2}',\lambda_B,l_{1}') I_\eta(x_A-x_B) 
G(m_{A1,B1}')
G(m_{A2,B2}')
\cr &\hphantom{ d^4k_1 }\times 
   \tspb{\smash{m_{A1}^{\prime} - x_A^{a}l'_{1\,a}}}.
        {\smash{m_{A2}^{\prime} - x_A^{a}l'_{2\,a}}}^{\e-2}
   \prod_{i\in A} \delta^2(\mu_{Ai}^{\adot} - x_A^{a\adot}\lambda_{Ai\,a})
   \delta^2(m_{B1}^{\prime\bdot} - x_B^{b\bdot}l'_{1\,b})
\cr &\hphantom{ d^4k_1 }\times 
   \prod_{i\in B} \delta^2(\mu_{Bi}^{\smash{\bdot}} 
                           - x_B^{b\smash{\bdot}}\lambda_{Bi\,b})
   \delta^2(m_{B2}^{\prime\bdot} - x_B^{b\bdot}l'_{2\,b}),
}\anoneqn$$
up to over-all $\e$-dependent factors with a smooth $\e\rightarrow0$ limit.
As noted in the previous section, it is straightforward to include 
fermion and scalar contributions.

\section{Dentistry with Differential Operators}
\tagsection\DifferentialOperatorSection
\vskip 10pt

Amplitudes in twistor space have simple properties.  At tree level,
they are supported, as it turns out, on loci of intersecting lines.
This means the amplitudes contain delta functions, so that the
twistor-space amplitudes satisfy certain algebraic relations.  In
particular, certain polynomials multiplying the amplitudes will yield
a function which vanishes everywhere.  Unfortunately, the coefficients
of the delta functions are quite difficult to calculate directly.

As Witten pointed out in his
original paper~[\use\WittenTopologicalString], however,
 we do not need the twistor-space amplitudes
in order to establish the structure of the delta functions they contain.
In momentum space, the Fourier transform turns the polynomials into
differential operators (polynomial in the $\lambda_i$, and derivatives
with respect to the $\tlambda_i$), which will annihilate the amplitude.
One particularly useful building block for these differential
operators is the line annihilation operator, expressing the condition
that three points in twistor space lie on a common `line' or $\CP^1$.
If the coordinates of the three points are 
$Z^I_{i=1,2,3}=(\lambda^a_i,\mu^{\adot}_i)$, the appropriate condition
is 
$$
\epsilon_{IJKL} Z^I_1 Z^J_2 Z^K_3 = 0,
\anoneqn$$
for all choices of $L$.  Choosing $L=\adot$, and translating this
equation back to momentum space using the identification 
$\mu^\adot \leftrightarrow -i\partial/\partial\lambda_\adot$, we 
obtain the operator,
$$
F_{123} = \tspa{\lambda_1}.{\lambda_2} 
{\partial\over\partial\tlambda_3}
+\tspa{\lambda_2}.{\lambda_3} 
{\partial\over\partial\tlambda_1}
+\tspa{\lambda_3}.{\lambda_1} 
{\partial\over\partial\tlambda_2}.
\anoneqn$$

The tree-level MHV amplitude, for example, is annihilated by $F_{ijk}$,
because it is independent of the $\tlambda_i$ (and any possible delta
functions vanish for generic momenta).  However, as noted by 
Cachazo, Svr\v cek, and Witten~[\use\CSWLoop], not all such
operators (nor products of them corresponding to configurations
with all points on two intersecting lines
[fig.~1(b) of that paper]) annihilate the finite parts of the 
one-loop amplitude.
This is why they originally interpreted parts of the amplitude as having
one of the points wandering off the line into the plane defined by their
intersection, and also why they interpreted the amplitude as having
derivative-of-delta function support in that plane.

If we interpret $\tlambda_i$ as the complex conjugate of $\lambda_i$,
as we must for real Minkowski momenta, then we must take into 
account~[\use\CSWLoopII],
the fact that $\partial_{\overline z}\, (1/z)\neq 0$; for spinor
products,
\def\lambdab{{\overline\lambda}}
\def\etab{{\overline\eta}}
\def\chib{{\overline\chi}}
$$
\eta^{\adot} {\partial\over\partial\lambdab^\adot}
{1\over \tspa{\lambda}.{\chi}} = 2\pi  \tspb{\etab}.{\chib}
 \delta(\tspa{\lambda}.{\chi})\delta(\tspb{\lambdab}.{\chib}),
\eqn\SpinorDeltaFunction$$
where $\lambdab_i$ denotes the complex conjugate of $\lambda_i$.
In integrating over the cut loop momenta to compute a one-loop amplitude,
we will necessarily encounter configurations where the delta function
does not vanish.  These give rise to the `holomorphic anomaly' contribution,
which we now proceed to compute explicitly for the MHV amplitude.

\def\anomaly{\Delta}
For notational clarity, let us take $c_1=1$, so that
we are considering the cut in the $s_{1\cdots c_2}$ channel,
$$\eqalign{
&\int {d^4\ell_1 \delta^{(+)}(\ell_1^2)\over (2\pi)^3}
{d^4\ell_2 \delta^{(+)}(\ell_2^2)\over (2\pi)^3}\;
(2\pi)^4 \delta^4(K_{1\cdots c_2}-\ell_1-\ell_2)\,
\cr &\hskip 10mm\times
A^\tree((-\ell_1)^+,1^+,\ldots,m_1^-,\ldots,m_2^-,\ldots,c_2^+,(-\ell_2)^+)
A^\tree(\ell_2^-,(c_2\!+\!1)^+,\ldots,n^+,\ell_1^-)
\cr &= C \int {d^4\ell_1 \delta^{(+)}(\ell_1^2)\over (2\pi)^3}
{d^4\ell_2 \delta^{(+)}(\ell_2^2)\over (2\pi)^3}\;
(2\pi)^4 \, \delta^4(K_{1\cdots c_2}-\ell_1-\ell_2)\,
{\tspa{\ell_1}.{\ell_2}^2\over\tspa{\ell_1}.{\lambda_1}
 \tspa{\lambda_{c_2}}.{\ell_2}
 \tspa{\ell_2}.{\lambda_{c_2+1}}\tspa{\lambda_n}.{\ell_1}},
}\anoneqn$$
where $\delta^{(+)}(\ell^2) = \Theta(\ell^0)\delta(\ell^2)$, and
with 
$$
C = -i \tspa{\lambda_{c_2}}.{\lambda_{c_2+1}}\tspa{\lambda_n}.{\lambda_1}
A^\tree(1^+,\ldots,m_1^-,\ldots,m_2^-,\ldots,n^+).
\anoneqn$$

Consider the result of applying $\eta^\adot F_{123\adot}$ 
to this cut.  Only the derivative
$\partial/\partial\lambdab_1$ can lead to a delta function term, because
only $\lambda_1$ amongst $\lambda_{1,2,3}$
appears in an inverse spinor product with one of the cut loop momenta $\ell_i$.
Using eq.~(\use\SpinorDeltaFunction), this delta function or `anomaly' term is,
\def\ellb{{\overline\ell}}
$$\eqalign{
\eta^{\adot}\anomaly_{123\adot} &= 
2\pi C 
\tspa{\lambda_2}.{\lambda_3} \int {d^4\ell_1 \delta^{(+)}
(\ell_1^2)\over (2\pi)^3}
{d^4\ell_2 \delta^{(+)}(\ell_2^2)\over (2\pi)^3}\;
(2 \pi)^4 \, \delta^4(K_{1\cdots c_2}-\ell_1-\ell_2)\,
\cr &\hskip 40mm\times
{\tspb{\etab}.{\ellb_1}\delta(\tspa{\lambda_1}.{\ell_1})
\delta(\tspb{\lambdab_1}.{\ellb_1})
\tspa{\ell_1}.{\ell_2}^2\over
 \tspa{\lambda_{c_2}}.{\ell_2}
 \tspa{\ell_2}.{\lambda_{c_2+1}}\tspa{\lambda_n}.{\ell_1}},
}\anoneqn$$

The cut integral, which has eight integration variables, also contains
eight delta functions.
As we shall show in Appendix~\use\HolomorphicAppendix, these are all
independent, and determine the integration variables completely.  The
delta functions require that $\tspa{\ell_1}.{\lambda_1}$ vanish, 
or in other words that
 $\ell_1$ be collinear to $k_1$,
$$
\ell_1 = a k_1.
\eqn\DeltaContraintA$$
The constant $a$ may be determined from the requirement that $\ell_2$
be massless,
$$
a = {s_{1\cdots c_2}\over 2 k_1\cdot K_{1\cdots c_2}},
\anoneqn$$
so that
\def\projn{\natural}
$$
\ell_2 = K_{1\cdots c_2} - {s_{1\cdots c_2}\over 
                 2 k_1\cdot K_{1\cdots c_2}} k_1 \equiv K^\projn
\;\hbox{\rm\ (`$K$-natural')}.
\eqn\DeltaContraintB$$

Because all the integration variables are determined, the result
is obtained by inserting the delta-function constraints
(\use\DeltaContraintA) and (\use\DeltaContraintB) into the 
product of tree amplitudes and multiplying by a 
jacobian factor, 
$$
2\pi C \, {\tspa{\lambda_2}.{\lambda_3} } \, {\cal J} \,
  {s_{1\cdots c_2}\over 2 k_1\cdot K_{1\cdots c_2}} \, 
{\tspb{\etab}.{\lambdab_1}
\tspa{\lambda_1}.{K^\projn}^2\over
 \tspa{\lambda_{c_2}}.{K^\projn}
 \tspa{K^\projn}.{\lambda_{c_2+1}}\tspa{\lambda_n}.{\lambda_1}}.
\eqn\AnomalyA
$$
The jacobian factor ${\cal J}$ arises from the integration 
over the delta functions and is computed in 
appendix~\use\HolomorphicAppendix.  The result is 
$$
{\cal J} = {\cg \over 2 k_1\cdot K_{1\cdots c_2}}.
\anoneqn$$
where the ${\cal O}(\e)$ terms implicit in $\cg$ are irrelevant 
because we are considering infrared finite amplitudes here.

\def\tsand#1.#2.#3{\sand{\lambda_{#1}}.#2.{\lambdab_{#3}}}
Multiplying and dividing eq.~(\use\AnomalyA) by
$\tspb{K^\projn}.{\lambdab_1}^2$, we can rewrite it without the
explicit appearance of $K^\projn$,
$$
\eta^{\adot}\anomaly_{123\adot} = 
-2\pi C \tspa{\lambda_2}.{\lambda_3}
{\tspb{\etab}.{\lambdab_1} s_{1\cdots c_2}\over
 \sand{\lambda_{c_2}}.{\Ksl_{1\cdots c_2}}.{\lambdab_1}
 \sand{\lambda_{c_2+1}}.{\Ksl_{1\cdots c_2}}.{\lambdab_1}
 \tspa{\lambda_n}.{\lambda_1}},
\anoneqn$$
where
$$
\sand{\lambda_1}.{\Ksl}.{\lambdab_2} 
= \lambda_{1}^a K_{a\adot} \smash{\lambdab_2}\vphantom{\lambda_2}^{\adot}.
\anoneqn$$
Although we have focused on the computation of the anomaly for the cut MHV
amplitude, the computation for other amplitudes is similar.  The jacobian
is universal; and the spinorial delta functions depend only on the
potential collinear singularities of the cut momenta with an external
momentum.  That in turn is governed by universal splitting amplitudes.

We can now define a modified line operator,
$$
\eta^{\adot}F_{123\adot}' = \eta^{\adot} 
\bigl(F_{123\adot} - \Delta_{123\adot}\bigr).
\anoneqn$$
The twistor-space picture expected from the BST computation, and
made explicit in sect.~\use\InfraredFiniteSection, is that 
the diagram corresponding to the cut considered here will
be supported on configurations where points $1,\ldots, c_2$ lie on
one line, and points $(c_2+1),\ldots,n$ lie on another line.
Taking into account the `holomorphic anomaly' means that this should be
reflected not in $F_{123} A_{\rm cut} = 0$, but rather in
$$
F'_{123} A_{\rm cut} = 0.
\anoneqn$$

\def\Cut{D}
Let us now check this equation.  
Using the spinor factorization~[\use\ZFourPartons]
$$
(P+p)^2 (P+q)^2 - P^2 Q^2 = \sand{p}.P.q \sand{q}.P.p,
\anoneqn$$
we can rewrite the cut~(\use\InfraredFiniteCut) in a more convenient
form (setting $c_1=1$),
$$\eqalign{
\Cut &=
2\pi i \cg A^\tree \cr
&\hphantom{=!}\hskip 2mm\times\ln \biggl({\tsand{n}.{\Ksl_{1\cdots c_2}}.{c_2}
                    \tsand{c_2}.{\Ksl_{1\cdots c_2}}.{n}
                   \tsand{1}.{\Ksl_{1\cdots c_2}}.{c_2+1}
                    \tsand{c_2+1}.{\Ksl_{1\cdots c_2}}.{1}
                   \over \tsand{1}.{\Ksl_{1\cdots c_2}}.{c_2}
                          \tsand{c_2}.{\Ksl_{1\cdots c_2}}.{1}
                          \tsand{n}.{\Ksl_{1\cdots c_2}}.{c_2+1}
                          \tsand{c_2+1}.{\Ksl_{1\cdots c_2}}.{n}
                  }\biggr).
}\eqn\CutRewritten$$
Now, $F'$ is a linear operator, so we can consider its action on
the cut as the sum of its actions on different factors inside the
cut.  In particular, $F_{123}$ (and hence $F'_{123}$) annihilates
any function of $K_{123}$, and thus will give no contribution from
acting on the $\Ksl_{1\cdots c_2}$ inside the cut~(\use\CutRewritten).
The only action left is that on the lone $\lambdab_1$s in $\Cut$,
that is, on
$$
2\pi i \cg A^\tree \ln \biggl({
                    \tsand{c_2+1}.{\Ksl_{1\cdots c_2}}.{1}
                   \over  \tsand{c_2}.{\Ksl_{1\cdots c_2}}.{1}
                  }\biggr);
\anoneqn$$
 computing that, we find
$$
\eta^{\adot} F_{123\adot} \Cut = 2\pi i \tspa{\lambda_2}.{\lambda_3} 
  \cg A^\tree
\biggl[  {\sand{\lambda_{c_2+1}}.{\Ksl_{1\cdots c_2}}.{\etab}
          \over\tsand{c_2+1}.{\Ksl_{1\cdots c_2}}.{1}}
        -{\sand{\lambda_{c_2}}.{\Ksl_{1\cdots c_2}}.{\etab}
          \over\tsand{c_2}.{\Ksl_{1\cdots c_2}}.{1}}
\biggr].
\anoneqn$$
Using the Schouten identity, we can simplify this to
$$
2\pi i\tspa{\lambda_2}.{\lambda_3}
 \cg A^\tree {s_{1\cdots c_2}\tspb{\etab}.{\lambdab_1}
                        \tspa{\lambda_{c_2}}.{\lambda_{c_2+1}}
                        \over \tsand{c_2}.{\Ksl_{1\cdots c_2}}.{1}
                              \tsand{c_2+1}.{\Ksl_{1\cdots c_2}}.{1}},
\anoneqn$$
so that 
$$
F'_{123} \Cut = 0.
\anoneqn$$
Cut constructibility then implies that $F'_{123}$ also
annihilates the dispersive parts of this diagram.

\section{Conclusions}
\vskip 10pt

Tree-level amplitudes in twistor space have a simple structure.  They
can be obtained from correlation functions evaluated on disconnected
$D$-instantons of degree one, linked by propagators.  Starting from
the BST construction of one-loop amplitudes based on MHV vertices, we
have shown that one-loop MHV amplitudes have a similar simple
structure.  They can also be obtained from correlation functions
evaluated on disconnected degree-one instantons, in this case linked
by two propagators.  This suggests that more general one-loop
amplitudes should have a similar structure.  In particular, the
discussion here, along with earlier predictions by
Dixon~[\use\NMHVPrivate] and also the picture emerging from NMHV loop
calculations by Dixon, Del Duca and two of the
authors~[\use\SevenPoint], suggests that the one-loop next-to-MHV
amplitudes should be obtained from the configurations shown in
fig.~\use\TwistorNMHVLoopFigure.  We have considered the ${\cal N}=4$
supersymmetric gauge theory in this paper, but it is clear that the
arguments apply equally well to any supersymmetric theory, because the
four-dimensional cuts suffice to construct the full
amplitude~[\use\NeqOneOneLoop].

\topinsert\LoadFigure\TwistorNMHVLoopFigure
{\baselineskip 13 pt
\noindent\narrower
The twistor-space configurations expected to contribute to a one-loop NMHV
amplitude, in accordance with earlier predictions by Dixon~
[\use\NMHVPrivate] and as motivated by NMHV one-loop calculations by Dixon,
Del Duca and two of the authors [\use\SevenPoint].
}  {\epsfxsize 4 truein}{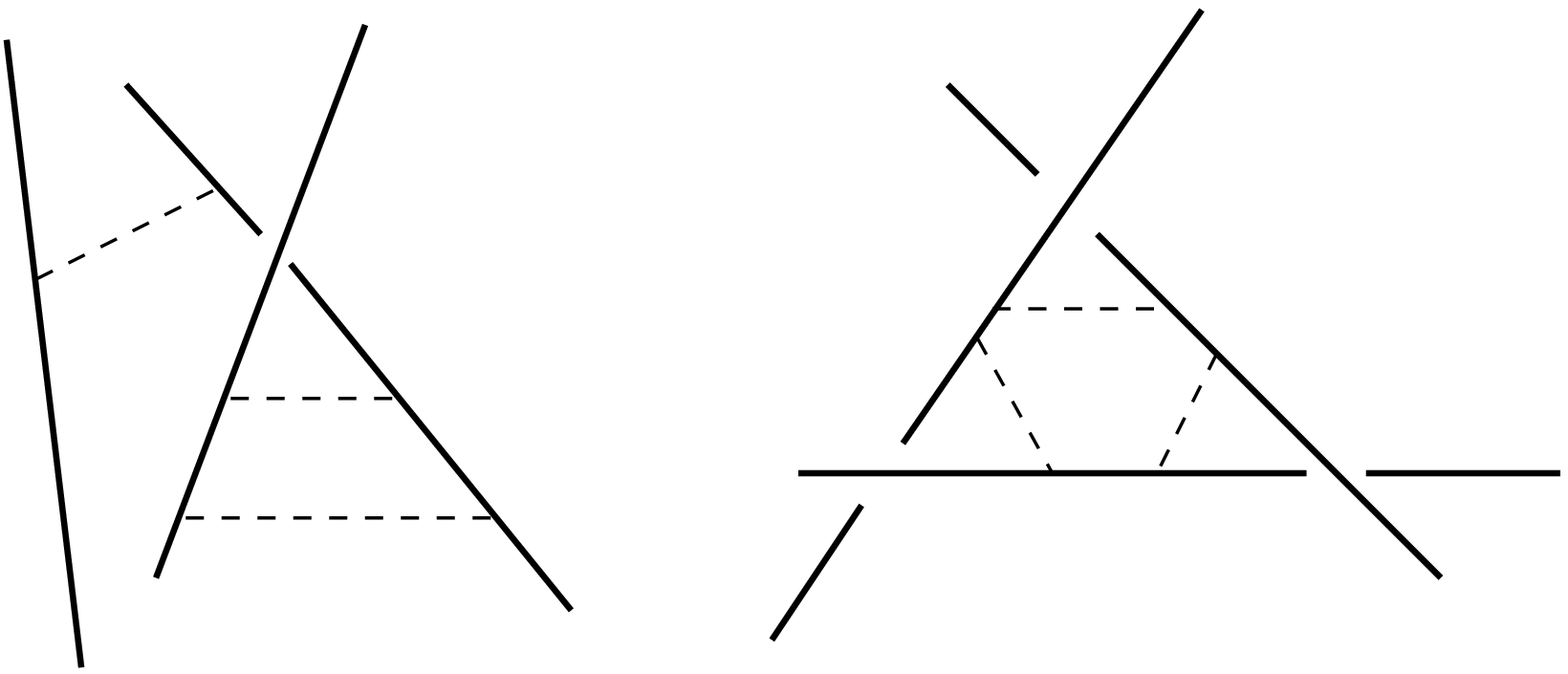}{}\endinsert

We have also shown that the infrared divergences characteristic of
loop amplitudes in gauge theories are isolated in a certain class
of twistor-space diagrams.  This means that the question of
how to derive generic diagrams from a twistor-space string theory
can be separated from the issue of infrared divergences.  We have
also shown how to introduce an infrared regulator corresponding
to a momentum-space dimensional regulator for the subset of diagrams
that do have infrared divergences.

We have evaluated the CSW `holomorphic anomaly' explicitly, and
shown that it can be used to define modified differential operators
whose kernels contain the one-loop MHV amplitudes in agreement
with the structure outlined above.

One may wonder to what extent an equivalence can exist between this
twistor-space picture and one in which the one-loop amplitudes are
supported on a lone higher-degree genus-one curve.  At the very least,
one would need a method of projecting out the contributions of the
superconformal gravitons to the amplitude. In the 
translation of the BST construction we have performed in this
paper, this projection is implemented by
restricting the propagators to be gauge-multiplet propagators.

\vskip .3 cm 

We thank L.~J.~Dixon and V.~P.~Nair for useful conversations, and the Kavli
Institute for Theoretical Physics, where this work was carried out,
for its generous hospitality.  This research was supported in part by
the National Science Foundation under grants PHY99--07949,
PHY00--98395, PHY00--99590 and PHY01--40151 and by the US Department of
Energy under grants DE--FG02--91ER40671, DE--FG02--91ER40618 and
DE--FG03--91ER40662.

\appendix{Fourier Transforming for Dimensional Regularization}
\tagappendix\FourierAppendix
\vskip10pt

In order to introduce an appropriate dimensional regulator for
infrared-divergent diagrams, we need to evaluate the Fourier transform
of $\tspb{\smash\tlambda_a}.{\smash\tlambda_b}^{-\e}$, which we do
in this appendix.

The required transform is,
$$\eqalign{ 
 & \int  {d^2\tlambda_{a,b}\over (2\pi)^4}~~ 
(\tlambda_a^{\dot1} \tlambda_b^{\dot2} 
 - \tlambda_a^{\dot2} \tlambda_b^{\dot1}- i \eta')^{-\e}
 e^{i( \mu_a^{\dot1} \tlambda_a^{\dot2} - \mu_a^{\dot2} \tlambda_a^{\dot1} 
       + \mu_b^{\dot1} \tlambda_b^{\dot2} - \mu_b^{\dot2} \tlambda_b^{\dot1})} 
\cr & = \int  {d^2\tlambda_{a,b}\over (2\pi)^4}
   (-i \tlambda_b^{\dot2} )^{- \e}  
  (i \tlambda_a^{\dot1} - i  {\tlambda_a^{\dot2} \tlambda_b^{\dot1} \over 
                        \tlambda_b^{\dot2}} + \eta)^{- \e}  
e^{i( \mu_a^{\dot1} \tlambda_a^{\dot2} - \mu_a^{\dot2} \tlambda_a^{\dot1} 
     + \mu_b^{\dot1} \tlambda_b^{\dot2} - \mu_b^{\dot2} \tlambda_b^{\dot1})},
}\anoneqn$$
where $\eta' = \eta \tlambda_b^{\dot2} $ is a regulator.

We use the formula (eq.~(3.2.4) of ref.~[\use\Erdelyi])
$$
\int (a+i x )^{- \e} e^{-i k x} = 
-{2 \pi \over \Gamma(\e)} (- k)^{\e -1} e^{a k },
\eqn\ErdelyiFormula$$
to do the $\tlambda_a^{\dot1}$ integral,
$$-\int {d^2\tlambda_{b} d\tlambda_a^{\dot2}\over (2\pi)^4}
{2 \pi \over \Gamma(\e)} (-i \tlambda_b^{\dot2} )^{- \e} e^{i(
\mu_a^{\dot1} \tlambda_a^{\dot2} + \mu_b^{\dot1} \tlambda_b^{\dot2} - \mu_b^{\dot2} \tlambda_b^{\dot1})}
(- \mu_a^{\dot2})^{\e-1} 
 \exp\Bigl[ -i \mu_a^{\dot2} {\tlambda_a^{\dot2} \tlambda_b^{\dot1} \over\tlambda_b^{\dot2}} 
     + \mu_a^{\dot2} \eta \Bigr].
\anoneqn$$
Next, do the $\tlambda_a^{\dot2}$ integral,
$$\eqalign{
&-{\e\over (2\pi)^2\,\Gamma(1+\e)} \int d^2\tlambda_{b}
(-i \tlambda_b^{\dot2} )^{-
\e} e^{i( \mu_b^{\dot1} \tlambda_b^{\dot2} - \mu_b^{\dot2} \tlambda_b^{\dot1})} (-
\mu_a^{\dot2})^{\e-1} \exp[ \mu_a^{\dot2} \eta ] \delta( \mu_a^{\dot1} - {\mu_a^{\dot2}
\tlambda_b^{\dot1} \over \tlambda_b^{\dot2}} )
\cr &=  -{\e\over (2\pi)^2\Gamma(1+\e)} 
\int  d\tlambda_{b}^{\dot2}
(-i \tlambda_b^{\dot2} )^{-\e}  (- \mu_a^{\dot2})^{\e-1} 
 \exp[ \mu_a^{\dot2} \eta'/\tlambda_b^{\dot2}]   (- \mu_a^{\dot2}   )^{-1} (-\tlambda_b^{\dot2}) 
\exp\Bigl[{i(\mu_b^{\dot1} \tlambda_b^{\dot2} 
        - {\mu_b^{\dot2} \mu_a^{\dot1} 
\tlambda_b^{\dot2} \over \mu_a^{\dot2}} )}\Bigr].
}\anoneqn$$ 
Only the sign of the coefficient of $\eta$ inside the exponential matters, so 
we can rewrite it as 
 $\exp[ \eta'\tlambda_b^{\dot2}/ \mu_a^{\dot2} ] $.  To do the last Fourier
transform we use the inverse of eq.~(\use\ErdelyiFormula), 
with $\e' = 2-\e$, obtaining 
$$\eqalign{& (i)^{- \e}  { \e\Gamma(2-\e) \over (2\pi)^3\Gamma(1+\e)}  
(- \mu_a^{\dot2}   )^{\e-2}
(  {i(  \mu_b^{\dot1} - {\mu_b^{\dot2} \mu_a^{\dot1} \over \mu_a^{\dot2}} )} + \eta'/ \mu_a^{\dot2} )^{\e-2} 
\cr & =  { \e\Gamma(2-\e) \over (2\pi)^3\Gamma(1+\e)} 
(\mu_b^{\dot2} \mu_a^{\dot1} -  \mu_b^{\dot1} \mu_a^{\dot2} + i \eta' )^{\e-2} 
\cr &=  { \e\Gamma(2-\e) \over  (2\pi)^3\Gamma(1+\e)} [ \mu_a,\mu_b]^{\e-2}.
}\anoneqn$$
As $\e \rightarrow 0$ this turns into a product of $\delta$ functions 
for all four components of the $\mu$s.
We can therefore regard it as defining a delta-convergent sequence, or
equivalently a smearing of the twistor-space delta function.


\appendix{Evaluation of the Jacobian}
\tagappendix\HolomorphicAppendix
\vskip 10pt

\def\la{\lambda}
\def\lb{{\overline \lambda}}

\newbox\charbox
\newbox\slabox
\def\s#1{{      
        \setbox\charbox=\hbox{$#1$}
        \setbox\slabox=\hbox{$/$}
        \dimen\charbox=\ht\slabox
        \advance\dimen\charbox by -\dp\slabox
        \advance\dimen\charbox by -\ht\charbox
        \advance\dimen\charbox by \dp\charbox
        \divide\dimen\charbox by 2
        \raise-\dimen\charbox\hbox to \wd\charbox{\hss/\hss}
        \llap{$#1$}
}}

In this appendix we compute the jacobian appearing in 
eq.~(\use\AnomalyA), arising from the
phase space integration over the anomaly-induced delta functions in 
eq.~(\use\SpinorDeltaFunction).  The jacobian is 
$$
{\cal J}= \int {d^4\ell_1 \delta^{(+)}(\ell_1^2)\over (2\pi)^3}
{d^4\ell_2 \delta^{(+)}(\ell_2^2)\over (2\pi)^3}\;
(2\pi)^4 \delta^4(K_{1\cdots c_2}-\ell_1-\ell_2)\, 
 \delta(\tspa{\lambda_{\ell_1}}.{\lambda_{\ell_2}})\,
\delta(\tspb{\lambdab_{\ell_1}}.{\lambdab_{\ell_2}}).
\anoneqn
$$
Expressing the phase space integral in terms of the CSW measure given in
eq.~({\use\CSWMeaure}) yields,
$$
\eqalign{
{\cal J}=  \cg & \int_0^\infty d T_1 \, \int_0^\infty d T_2 \, 
\int_{\C} d w_1  d \overline w_1 \int_{\C} d w_2 d \overline w_2 \;  T_2 \,
  \delta(\la_1^2 - \la_1^1 w_1) \,
      \delta(\lb_1^{\dot 2} - \lb_1^{\dot 1} \overline w_1)\cr
& \times 
\delta^4 \! \left( 
  \left(\eqalign{K_{1\cdots c_2}^0 + K_{1\cdots c_2}^3 
             \hskip .4 cm & K_{1\cdots c_2}^1 - i K_{1\cdots c_2}^2 \cr
          K_{1\cdots c_2}^1 + i K_{1\cdots c_2}^2 \hskip .3cm & \hskip .1 cm 
          K_{1\cdots c_2}^0 -  K_{1\cdots c_2}^3 \cr } \right)
 - T_1 
\left( \eqalign{ 1 \hskip .4 cm    & \hskip .3 cm \overline w_1 \cr
      w_1 \hskip .2 cm   &  \hskip .1 cm w_1 \overline w_1 } \right)  
 - T_2
\left( \eqalign{ 1 \hskip .4 cm    & \hskip .3 cm \overline w_2 \cr
      w_2 \hskip .2 cm   &  \hskip .1 cm w_2 \overline w_2 } \right)  
\right), \cr 
 }
\anoneqn
$$
where we have set $T_1 = \tau_1^2$ and $T_2 = \tau_2^2$.  (In
this expression, $\e$ in the $\cg$ factor is set to zero since here we
consider infrared finite diagrams.)
The $\delta$-functions yield a set of equations, 
$$
\eqalign{
& f_1 \equiv  \la_1^2  - \la_1^1 w_1 = 0, \cr
& f_2 \equiv \lb_1^{\dot 2}  - \overline w_1 \lb_1^{\dot 1} = 0 , \cr
& f_3 \equiv
 - T_2 w_2 -T_1 w_1 + ( K_{1\cdots c_2}^1 + i  K_{1\cdots c_2}^2) =0,\cr
& f_4 \equiv - T_2 \overline w_2 - T_1 \overline w_1 + ( K_{1\cdots c_2}^1 
                            - i  K_{1\cdots c_2}^2) =0 , \cr
& f_5 \equiv -T_1 - T_2 + ( K_{1\cdots c_2}^0 +  K_{1\cdots c_2}^3) = 0, \cr
& f_6 \equiv -T_1 w_1 \overline w_1 - T_2 w_2 \overline w_2 
  + (K_{1\cdots c_2}^0 - K_{1\cdots c_2}^3)   = 0 . \hskip 5 cm  }
\eqn\DeltaEquations
$$
These equation are all independent and hence fix all integration
variables.  Integrating the over the delta function leads to the
jacobian
$$
\eqalign{
{\cal J} & = \cg T_2 \Bigl| {\partial(f_1, f_2, f_3, f_4, f_5, f_6) \over
                \partial(w_1, w_2, \overline w_1, \overline w_2 ,T_1, T_2)}
              \Bigr|^{-1} \cr
& = \cg \Bigl[ {T_2 \lambda_1^1 \overline \lambda_1^{\dot 1}
     (w_2 - w_1) (\overline w_2 - \overline w_1) } \Bigr]^{-1} \cr
& = \cg {1\over 2 k_1 \cdot K_{1\cdots c_2}}, \cr}
\anoneqn
$$
where we substituted the solution to the relations ({\use\DeltaEquations})
to obtain the result on the last line.
This jacobian factor is independent of the particular one-loop cut amplitude
under consideration and applies just as well to the non-MHV case.

\listrefs
\bye